\documentclass[a4paper,11pt]{article}

\usepackage{jcappub}
\usepackage{etoolbox}
\makeatletter
\newcommand{\preprint}[1]{\gdef\@mypreprint{#1}}
\newcommand{\@mypreprint}{}
\pretocmd{\maketitle}{%
  \ifx\@mypreprint\@empty\else
    \begin{flushright}\@mypreprint\end{flushright}\vspace{-1.2em}%
  \fi
}{}{}
\makeatother

\preprint{CETUP2025-018}
\usepackage[utf8]{inputenc}
\usepackage[T1]{fontenc}

\usepackage[mathscr]{eucal}
\usepackage[Symbolsmallscale]{upgreek}
\usepackage{xcolor}
\usepackage{slashed}
\usepackage{hyperref}

\usepackage[compat=1.1.0]{tikz-feynman}
\usepackage{tikz}

\usepackage{pdfpages}

\definecolor{darkgreen}{rgb}{0.0,0.5,0.0}

\title{
 Gravitational Waves from Multiple First-Order Phase Transitions in a Scenario with Early Matter Domination
}

\author[a]{Rouzbeh Allahverdi,}
\author[b]{Fazlollah Hajkarim}

\affiliation[a]{Department of Physics \& Astronomy, University of New Mexico, Albuquerque, NM 87106, USA}
\affiliation[b]{Department of Physics and Astronomy, University of Oklahoma, Norman, OK 73019, USA}

\emailAdd{rouzbeh@unm.edu}
\emailAdd{fazlollah.hajkarim@ou.edu}

\abstract{
Non-standard cosmological histories with epochs of early matter domination (EMD) arise in various top-down models of the early universe. Typically, in the latter stage of EMD, temperature decreases more slowly than in a radiation dominated universe because of entropy generation from decay of the species that drives EMD. A time-dependent decay rate can significantly modify this picture and even lead to a period with increasing temperature. We study non-monotonic temperature evolution in a well-motivated scenario of EMD with a time-dependent decay rate that can give rise to multiple first-order phase transitions in both cooling and heating phases. The spectra of the ensuing gravitational waves (GW) exhibit characteristic features such as multiple peaks and a distinct behavior at high frequencies. These features allow us to determine the phase transition temperature as well as the reheating temperature at the end of the EMD. The future GW detectors can therefore provide a probe for the new physics and a window to the early thermal history.
}

\begin{document}

\maketitle
\flushbottom


\section{Introduction}
\label{sec:intro}
The measurement of gravitational waves (GW) from binary mergers by the LIGO and Virgo collaborations \cite{LIGOScientific:2016aoc} has confirmed one of the main predictions of general relativity. 
In addition to GW produced from compact binaries, we generally expect a stochastic GW background of cosmological origin. This component could arise from first-order phase transitions (FOPTs) \cite{Witten:1984rs,Hindmarsh:2017gnf,Weir:2017wfa}, cosmic string networks \cite{Damour:2004kw}, and formation of primordial black holes (PBHs)  \cite{Carr:1975qj,Carr:2020xqk,Carr:1974nx,Green:2020jor} in the early universe. Detecting GW from such events can reveal important information about the early universe \cite{Christensen:2018iqi}. Near future GW detectors such as LISA \cite{Audley:2017drz}, the Einstein Telescope (ET) \cite{Punturo:2010zz}, and Cosmic Explorer (CE) \cite{Reitze:2019iox} can reach the sensitivity required to detect signals of primordial origin and thereby open a probe of new physics beyond the terrestrial experiments \cite{Ramsey-Musolf:2019lsf}. 

GW also opens a window to the very early universe for which we currently have no direct observational probe. In particular, the thermal evolution of the universe before big bang nucleosynthesis (BBN) remains an open question. In the standard picture, the universe entered a radiation dominated (RD) phase shortly after inflation and stayed there until matter-radiation equality. However, it is plausible that the post inflationary universe went through a non-standard expansion history (for reviews on and consequences of such histories, see~\cite{Allahverdi:2020bys,Batell:2024dsi}). 
Several scenarios describing the early Universe (especially those inspired by string theory~\cite{Kane:2015jia}) typically predict non-standard thermal histories during which one or more epochs of early matter domination (EMD) can happen. 
An EMD phase arises when a non-relativistic component temporarily becomes the dominant contributor to the energy budget of the Universe before decaying and restoring RD prior to the onset of BBN.
This behavior can originate from the coherent oscillatory evolution of a scalar degree of freedom, for instance a string modulus, which is displaced during inflation from the minimum of its potential~\cite{Coughlan:1983ci,deCarlos:1993wie,Banks:1993en,Allahverdi:2013noa,Felder:1998vq}, or from a long-lived field or particle produced after the inflationary era~\cite{Berlin:2016gtr,Dror:2017gjq,Cirelli:2018iax,Allahverdi:2021grt}.  

Cosmological FOPTs appear in many scenarios beyond the standard model (SM) of particle physics. They proceed via bubble nucleation from a false vacuum to a true vacuum \cite{Coleman:1977py,Linde:1981zj}~\footnote{FOPTs can also produce PBHs and dark matter (DM) \cite{Goncalves:2024vkj,Allahverdi:2024ofe}.
PBHs can contribute to both DM and GW signals from binary mergers producing a stochastic background detectable by current and future interferometers  \cite{Sasaki:2018dmp,Khlopov:1980mg,Crawford:1982yz,Hawking:1982ga,Chapline:1975ojl,Carr:2016drx,Bird:2016dcv,Clesse:2016vqa,Sasaki:2016jop,Ai:2024cka,Barni:2024lkj,Byrnes:2018clq,Domenech:2020ssp}.}. GW are sourced by collisions of these bubbles as well as produced sound waves and turbulence in the primordial plasma during PT~\cite{Caprini:2015zlo,Caprini:2019egz,Hindmarsh:2013xza,Hindmarsh:2015qta,Hajkarim:2025efr,Guo:2024gmu}. The peak frequency and amplitude of generated GW depend on the transition dynamics (which is governed by the underlying particle physics model) \cite{Grojean:2006bp,Huang:2016cjm}.  Reheating and non-standard thermal histories with EMD epochs affect GW production from FOPTs and DM production ~\cite{
Barenboim:2016mjm,Banik:2025olw,Buen-Abad:2023hex,Barman:2026kab,Allahverdi:2024ofe,Guo:2024kfk,Arias:2022qjt,Drees:2017iod,Drees:2018dsj,Cline:2026cek,Kane:2015jia,DEramo:2019tit,Allahverdi:2020bys,Conaci:2024tlc,Lee:2026fpg}.
In addition, because of different expansion histories, 
the spectrum of GW from PTs will be modified in scenarios with EMD \cite{Ellis:2020nnr,Guo:2020grp}.

In this paper, we focus on the dynamics of decay of the matter-like component that drives EMD and its implications for GW production from FOPTs. A time-dependent decay rate leads to a markedly different temperature evolution during EMD and can even lead to a period of increasing temperature towards the end of this epoch~\cite{Co:2020xaf}. This, for example, can happen if the matter-like component is identified with a flat direction in the scalar potential of supersymmetric (SUSY) extensions of the SM. The non-monotonic behavior of temperature can give rise to multiple PTs in both cooling and heating phases at the same temperature. We compute the spectrum of GW in such a scenario and show that it typically has multiple peaks accompanied by a shallow fall off $\propto f^{-1}$ at high frequencies.  These characteristic features can be used to determine the PT temperature as well as temperature at the end, and possibly the onset, of an EMD epoch with non-monotonic temperature evolution. We demonstrate this for benchmark points that lead to observable effects at frequencies within the reach of DECIGO, BBO, CE, and ET detectors~\cite{Bartolo:2016ami,Reitze:2019iox,Caprini:2015zlo,Baker:2019nia,Isoyama:2018rjb,Sathyaprakash:2012jk,ET:2019dnz}.

The rest of this paper is organized as follows. In Section~\ref{sec:bfeqs}, we discuss non-monotonic temperature evolution in an EMD scenario with time-varying decay rate and a well-motivated particle physics realization of it. In Section~\ref{sec:pt-gw}, we discuss production of GW from multiple FOPTs in both cooling and heating phases. We present our result for the spectrum of GW thus produced and its characteristic features in Section~\ref{sec:result}. We summarize and conclude the paper in Section~\ref{sec:conc-sum}. Some complementary details of calculations  are explained in the Appendices.  

\section{Evolution of Temperature During Early Matter Domination}
\label{sec:bfeqs}
We assume a period of EMD that begins in the RD universe, following the completion of the reheating process after the inflation. The matter component $\phi$ carries an energy density $\rho_\phi$ and decays at the rate $\Gamma_\phi$ to relativistic particles. The decay feeds the radiation energy density $\rho_{\rm R}$ and the subsequent evolution is governed by the following system of Boltzmann equations:
\begin{eqnarray} 
\label{Boltz1}
&& {d \rho_\phi \over dt} + 3 H \rho_\phi = - \Gamma_\phi \rho_\phi \,, \, \nonumber \\
&& {d \rho_{\rm R} \over dt} + 4 H \rho_{\rm R} = \Gamma_\phi \rho_\phi \, , 
\end{eqnarray}
where the Hubble expansion rate is set by the total energy density $\rho_\phi + \rho_{\rm R}$.  We have considered a more detailed set of Boltzmann-Friedmann equations in Appendix \ref{app:boltz-fried}.

Taking the onset of EMD as the moment when $\rho_\phi = \rho_{\rm R}$, we have:
\begin{equation}
\label{onset}
H_{\rm O} = \left({\pi^2 g_{*,{\rm O}} \over 45} \right)^{1/2} {T^2_{\rm O} \over M_{\rm P}} ,   
\end{equation}
where $g_{*,{\rm O}}$ is the number of relativistic degrees of freedom at temperature $T_{\rm O}$.

The EMD epoch consists of adiabatic and non-adiabatic phases (in the same chronological order). In the adiabatic phase, the initial radiation dominates over that produced from $\phi$ decay. Transition to the non-adiabatic phase occurs when radiation produced by $\phi$ decay becomes dominant. Eventually, $\phi$ decay completes resulting in a RD universe.

\begin{figure}[htbp!]
\centering



    \includegraphics[width=0.3\textwidth]{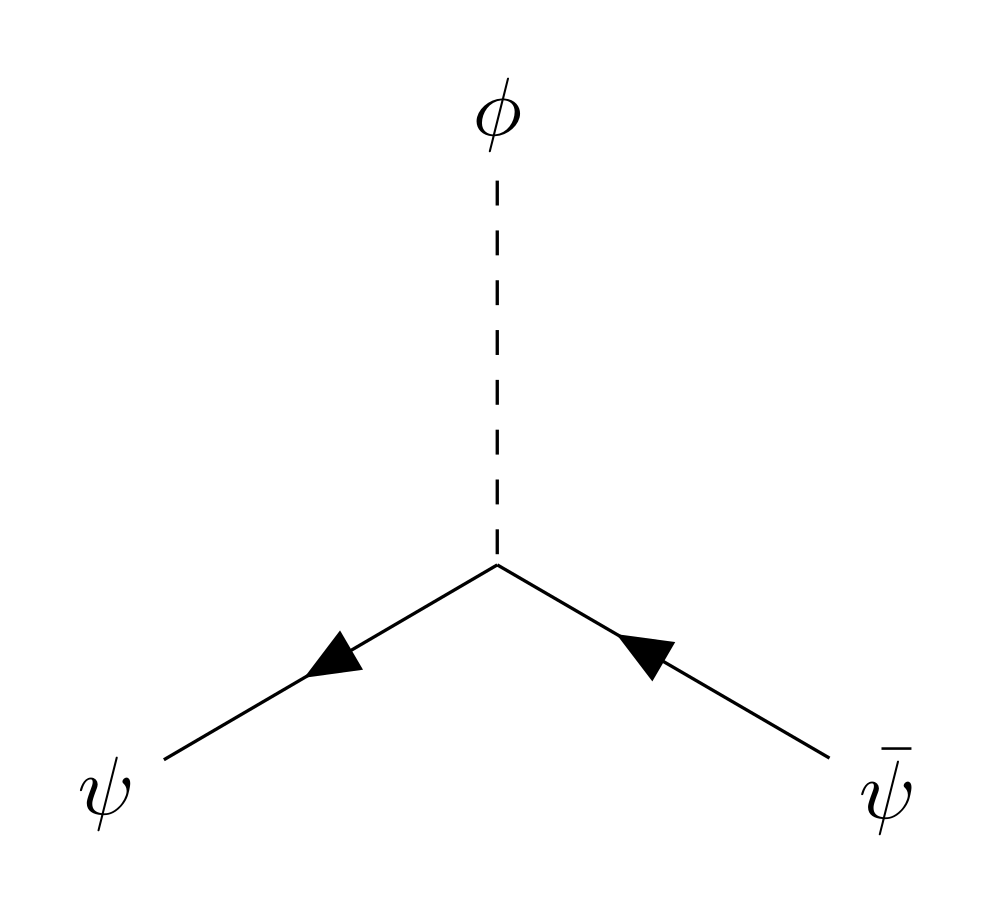}
    \hspace{0.10\textwidth}
    \includegraphics[width=0.3\textwidth]{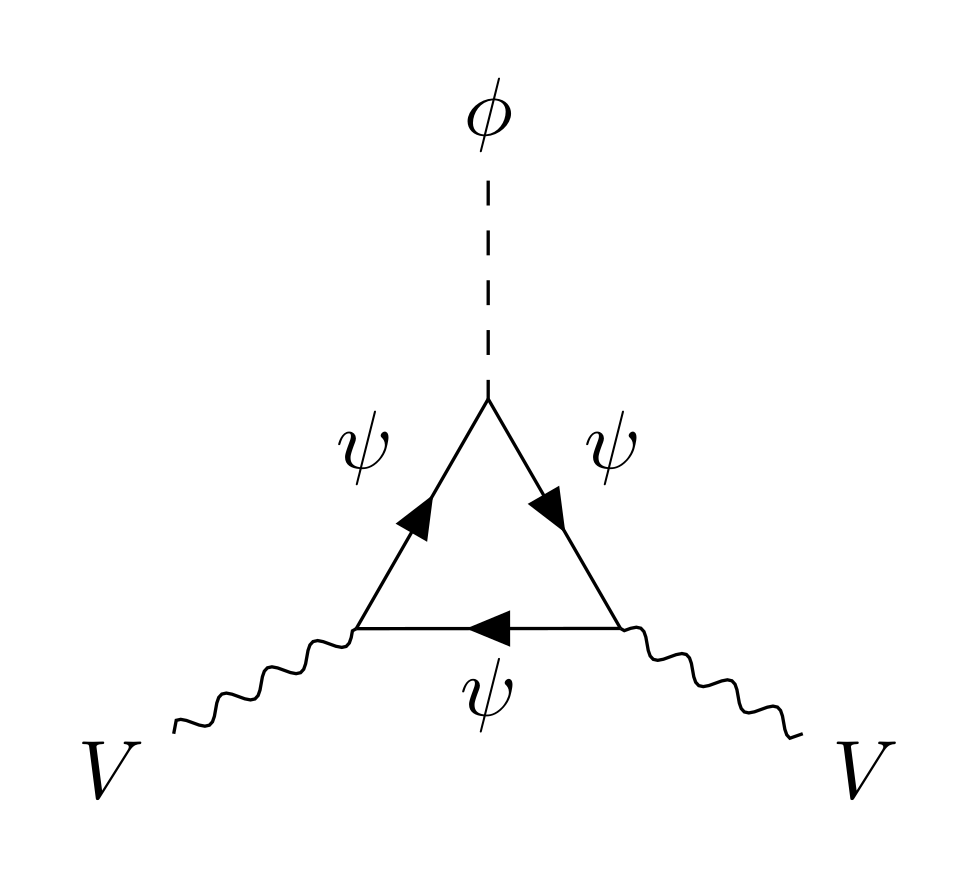}

\caption{
Left panel: tree-level diagram for $\phi$ decay to fermions. 
Right panel: one-loop diagram for $\phi$ decay to gauge bosons.
}
\label{fig:phi_decay}
\end{figure}

\subsection{The Standard Case}

This is the commonly studied case with $\Gamma_\phi$ being constant. The adiabatic phase of EMD, during which temperature is redshifted similar to $T \propto a^{-1}$, starts at $H_{\rm O}$. The relation between $H$ and $T$ in this phase follows:
\begin{equation} 
\label{ad}
H \simeq \left({\pi^2 g^{3/4}_{*} g^{1/4}_{*,{\rm O}} \over 90}\right)^{1/2} {T^{3/2} T^{1/2}_{\rm O} \over M_{\rm P}},
\end{equation}
where $g_{*}$ denotes the number of relativistic degrees of freedom at $T$.

This phase eventually ends when the redshifted value of the initial radiation becomes approximately equal to that produced from $\phi$ decay. The corresponding Hubble rate follows
\begin{equation}
\label{tr1}
H_{\rm EQ} \simeq 8^{1/5} H^{2/5}_{\rm O} \Gamma^{3/5}_\phi .
\end{equation}
From here on, the non-adiabatic phase of EMD begins, during which
\begin{equation} 
\label{nonad1}
H \simeq {\pi^2 g_{*} \over 90} ~ {T^4 \over \Gamma_\phi M^2_{\rm P}}.    
\end{equation}
Given that $H \propto a^{-3/2}$ during EMD, we see that
\begin{equation} 
\label{T1}
T \propto a^{-3/8} ~ ~ ~ ~ ~ ~ ({\rm non-adiabatic}) .
\end{equation}

The EMD epoch ends when $H \simeq \Gamma_\phi$ resulting in a RD universe with the following reheating temperature 
\begin{equation} 
\label{R1}
T_{\rm RH} \simeq \left({\pi^2 g_{*,{\rm RH}} \over 90}\right)^{-1/4} \left(\Gamma_\phi M_{\rm P}\right)^{1/2},
\end{equation}
where $g_{*, {\rm RH}}$ is the number of relativistic degrees of freedom at $T_{\rm RH}$.

\subsection{The Case with a Time-Varying Decay Rate}

In general, it is possible that $\Gamma_\phi$ depends on time. In a well-motivated case (see next subsection for more details) $\Gamma_\phi \propto \rho^{-1}_\phi$~\footnote{The impact of field-dependent decay rate during reheating dynamics on the production of DM and axion has been studied before \cite{Arias:2022qjt,Co:2020xaf,Laine:2010cq,Mukaida:2012qn}.}. Then, given that $\rho_\phi \propto a^{-3}$ and $a \propto H^{-2/3}$ during the EMD epoch, we can write 
\begin{equation} 
\Gamma_\phi(H) = \Gamma_{\phi,{\rm O}} {H^2_{\rm O} \over H^2},
\end{equation}
where $\Gamma_{\phi,{\rm O}}$ denotes the value of $\Gamma_\phi$ at the onset of EMD. 

The $H-T$ relation in the adiabatic phase remains the same as that in the standard case, see Eq.~(\ref{ad}). However, transition to the non-adiabatic phase now happens at
\begin{equation} 
\label{tr2}
H_{\rm EQ} \simeq 8^{1/11} \Gamma^{3/11}_{\phi ,{\rm O}} H^{8/11}_{\rm O}.     
\end{equation}
During the non-adiabatic phase of EMD we have
\begin{equation} 
\label{nonad2}
H \simeq \left({\pi^2 g_{*} \over 90}\right)^{-1} ~ {\Gamma_{\phi,{\rm O}} H^2_{\rm O} M^2_{\rm P} \over T^4}.    
\end{equation}
This results in a distinct behavior of temperature in the non-adiabatic phase
\begin{equation}
\label{T2}
T \propto a^{3/8} ~ ~ ~ ~ ({\rm non-adiabatic}) ,   
\end{equation}
which actually implies an increase of $T$.
The EMD epoch ends when $\Gamma_\phi(t) \simeq H$ leading to the following reheating temperature
\begin{equation} 
\label{R2}
T_{\rm RH} \simeq \left({\pi^2 g_{*,{\rm RH}} \over 90}\right)^{-1/4} \left(\Gamma_{\phi,{\rm O}} H^2_{\rm O} M^3_{\rm P}\right)^{1/6}.    
\end{equation}

\subsection{A Model with Time-Dependent Decay Rate}

It is well-known that the scalar potential of SUSY extensions of the SM have a large number of flat directions where $V = 0$ at the renormalizable level in the limit of unbroken SUSY~\cite{Gherghetta:1995dv}. These flat directions can acquire large vacuum expectation values during inflation. They start moving about the true minimum of the potential when the Hubble rate drops below their mass, which is set by soft SUSY breaking scale in the visible sector. For sufficiently large displacements, a flat direction can dominate the energy density of the universe thereby driving a period of EMD~\cite{Affleck:1984fy}. 

We consider a simple model, motivated by SUSY flat directions, where a complex scalar field $\phi$ of mass $m_\phi$ is coupled to massless fermions as follows: 
\begin{equation} 
\label{lagran1}
{\cal L} \supset - m^2_\phi \vert \phi \vert^2 + (i h ~ \phi ~ \psi^{\rm T}_{1} \sigma^2 \psi_{2} + {\rm h.c.}) .  
\end{equation}
Here $\psi_1$ and $\psi_2$ are two-component left-handed (LH) Weyl fermions and $h$ is a Yukawa coupling (taken to be a real positive number for simplicity). This model provides an accurate description of flat direction couplings to fermions~\footnote{Here we do not discuss couplings of the flat direction to the SUSY partners of fermions. But the consequence for $\phi$ decay will be similar.}. To put things in the context, $\phi$ could be identified by the $H_{\rm u} H_{\rm d}$ flat direction in which case $\psi_1$ and $\psi_2$ are the LH quarks/leptons and antiquarks/antileptons respectively.

By defining a right-handed (RH) Weyl spinor $\chi_1 = i \sigma^2 \psi^*_2$, one can form a four-component spinor $\Psi$ out of $\psi_1$ and $\chi_1$. Eq.~(\ref{lagran1}) can then be recast in the following form (we assume that $h$ is real)
\begin{equation} 
\label{lagran2}
{\cal L} \supset - m^2_\phi \vert \phi \vert^2 + h ~ \phi \Psi^{\dagger}_{\rm R} \Psi_{\rm L} + h ~ \phi^* \Psi^{\dagger}_{\rm L} \Psi_{\rm R} ,      
\end{equation}
where $\Psi_{L,R} = (1 \mp \gamma_5) \Psi/2$
are the LH and RH chiral components of $\Psi$.

A flat direction moves in both radial and angular directions where the latter is due to a torque exerted by the SUSY breaking $A$-term in the scalar potential~\cite{Dine:1995kz} (which is not shown above). This results in an elliptical trajectory in the $\phi$ plane where the eccentricity is typically much smaller than 1. Hence, for simplicity, we consider a circular trajectory given by
\begin{equation} 
\label{traj1}
\phi(t) = {\tilde \phi}(t) e^{i m_\phi t},    
\end{equation}
where ${\tilde \phi}(t)$ denotes the instantaneous radius of the circle. For $H \ll m_\phi$, the Hubble expansion slowly redshifts ${\tilde \phi}$ resulting in
\begin{equation} 
\label{traj2}
\rho_\phi (t) = m^2_\phi {\tilde \phi}^2(t) \,.
\end{equation}
Given that $\rho_\phi \propto a^{-3}$ during EMD, for $H_{\rm RH} < H < H_{\rm O}$, we have~\footnote{For an elliptical trajectory, there is another fast time-dependent component in ${\tilde \phi}(t)$ of frequency $m_\phi$. However, as long as the eccentricity is much smaller than 1, this rapid time variation does not lead to resonant particle production~\cite{Allahverdi:1999je}.}
\begin{equation} \label{traj3}
{\tilde \phi}(t) = {H \over H_{\rm O}} {\tilde \phi}_{\rm O}. 
\end{equation}

Plugging the expression in Eq.~(\ref{traj1}) into~(\ref{lagran2}), we have
\begin{equation} 
\label{lagran3}
{\cal L} \supset h {\tilde \phi}(t) e^{i m_\phi t} ~{\Psi}^{\dagger}_{\rm R} \Psi_{\rm L} + h {\tilde \phi}(t) e^{-i m_\phi t} ~ {\Psi}^{\dagger}_{\rm L} \Psi_{\rm R} .     
\end{equation}
The phase due to the rotational motion can be removed by a chiral phase transformation of $\Psi$ resulting in~\footnote{The $\Psi$ kinetic term gives rise to a term proportional to $m_\phi$ in ${\cal L}$ after this phase transformation. However, for $h {\tilde \phi} \gg m_\phi$, its effect is negligible relative to the $\Psi$ mass term.}
\begin{equation}
{\cal L} \supset h {\tilde \phi}(t){\bar \Psi} \Psi.
\end{equation}

We see that $\phi$ rotation induces a mass $h {\tilde \phi}(t)$ for $\Psi$, which kinematically blocks tree-level decay of $\phi$ to fermions, shown in the left panel of Fig.~\ref{fig:phi_decay}, as long as $h {\tilde \phi}(t) > m_\phi/2$. However, $\phi$ can decay at loop level to particles that are not directly coupled to it at the loop level. In particular, $\phi$ can decay to massless gauge fields via the fermion loop shown in the right panel of Fig.~\ref{fig:phi_decay}.

To demonstrate this, let us again consider the $H_{\rm u} H_{\rm d}$ flat direction. In this case, the nonzero value of ${\tilde \phi}$ spontaneously breaks $SU(2)_{\rm W} \times U(1)_{\rm Y}$ down to $U(1)_{\rm em}$. While the gauge fields of the broken subgroup acquire a mass $\sim g {\tilde \phi}$, with $g$ denoting a gauge coupling, the photon and gluons remain massless. As a result $\phi$ decays to a pair of photons (with leptons or quarks running in the loop) or gluons (with quarks in the loop). For $h {\tilde \phi} \gg m_\phi$, the decay rate is (see Appendix \ref{sec:appA} for details)
\begin{equation} 
\label{loopdec}
\Gamma_\phi^{\rm loop} (t) \simeq C \cdot {g^4 h^2 \over (4 \pi)^4 \cdot 8 \pi} \cdot {m^3_\phi \over m^2_\Psi} = C \cdot {g^4 \over (4 \pi)^4 \cdot 8 \pi} \cdot {m^3_\phi \over {\tilde \phi}^2(t)} ~ ~ ~ ~ ~ ~ 2 h {\tilde \phi} > m_\phi  ,  
\end{equation}
where $C$ is a multiplicity factor that accounts for the number of fermions in the loop and the number of final states. 

If $\Gamma_\phi^{\rm loop} > H$ when $2 h {\tilde \phi} > m_\phi$, then $\phi$ decay via the one-loop diagram will be efficient in terminating the EMD epoch. This happens to be the case, provided that
\begin{equation} 
\label{cond}
h > \left[{\sqrt{3} C g^4 \over 2^{11} \cdot \pi^5}\right]^{-1/3} \left({m_\phi \over M_{\rm P}}\right)^{1/3}. 
\end{equation}
Otherwise, the tree-level decay of $\phi$ to $\Psi$ opens up when $h {\tilde \phi}$ drops below $m_\phi/2$. For $h {\tilde \phi} \ll m_\phi$, the corresponding decay rate is 
\begin{equation} 
\label{eq:treedec}
\Gamma^{\rm tree}_{\phi} \simeq {h^2 \over 8 \pi} m_\phi   \,.
\end{equation}
In this case, the non-adiabatic phase of EMD switches to the standard case where $T \propto a^{-3/8}$ that ends when $\Gamma^{\rm tree}_{\phi} \gtrsim H$~\footnote{SUSY $D$-terms can in principle open non-perturbative decay channels of $\phi$ even for $h {\tilde \phi} \gg m_\phi$~\cite{Olive:2006uw}. However, this does not happen for the case with a single flat direction, or simplest cases with multiple flat directions~\cite{Allahverdi:2006xh}. Therefore, our discussion of $\phi$ decay remains intact in these cases.}.   

If the condition in Eq.~(\ref{cond}) is satisfied, then we will have a situation where $T \propto a^{-1}$ in the adiabatic phase of EMD is followed by $T \propto a^{3/8}$ in the non-adiabatic phase and then $T \propto a^{-1}$ during RD. This is the most interesting case for us as the fall and subsequent rise of $T$ can lead to PTs in the cooling and heating phases, respectively, during EMD. 

We have verified this picture by numerically solving the system of Boltzmann equations in~(\ref{Boltz1}) (see Appendix B for details). Fig. \ref{fig:m-h} depicts a scatter plot in the $m_\phi-h$ plane for $T_{\rm O} = 10^{11}$ GeV. The chosen range of $h$ is in agreement with the Yukawa couplings of the SM, and the values of $m_\phi$ satisfy $m_\phi > H_{\rm O}$ so that the $\phi$ field behaves like a matter component at the onset of EMD. 
In the entire range of $h$ and $m_\phi$ shown in the figure we have  $T_{\rm RH} \gtrsim 3$ MeV, which is required for the EMD epoch to end before the onset of BBN~\cite{deSalas:2015glj,Hasegawa:2019jsa}.

The blue dots correspond to cases where the loop diagram is responsible for $\phi$ decay and $T\propto a^{3/8}$ throughout the non-adiabatic phase of EMD epoch. The gray dots represent cases where the tree-level decay channel opens up during the non-adiabatic phase and becomes responsible for ending EMD. The red dots correspond to those cases where $2 h {\tilde \phi} < m_\phi$ already in the adiabatic phase of EMD. Then $\phi$ decay in the non-adiabatic phase is governed by the tree-level decay resulting in $T \propto a^{-3/8}$, and hence monotonic temperature evolution during EMD (i.e., the standard case).  

We see that in a significant fraction of the $m_\phi-h$ plane temperature evolution during EMD is non-monotonic and can lead to multiple PTs. We also see agreement between analytical estimates and numerical solutions, as the boundary between the red and gray dots follows the relation in Eq.~(\ref{cond})  for typical values of $g^2/4\pi \sim {\cal O}(10^{-2})$ and $C \sim {\cal O}(1)$.

The following remark is important at this point. We have ignored thermal effects on $\phi$ decay in our discussion. If $h {\tilde \phi} \gg T$, then $\Psi$ quanta will not be accessible to the thermal bath. On the other hand, when $h {\tilde \phi} \lesssim T$, quanta of $\Psi$ will be produced by the thermal bath and induce a thermal mass $\sim h T$ for the $\phi$ field ~\cite{Kapusta:2006pm,Drees:2021lbm,Drees:2022vvn,Co:2020xaf}. In any case, this contribution is negligible as long as $h~T < m_\phi$.  The gauge bosons also acquire a thermal mass $\sim g~T$. Thus, in order for the loop-level $\phi$ decay to proceed, we require that $m_\phi \gtrsim 2~ g~T$.  Thermal effects on $\phi$ decay can therefore be ignored provided that $h T < m_\phi$ and $m_\phi > 2~g~T$ \cite{Yokoyama:2004pf,Yokoyama:2006wt,Drees:2021lbm,Drees:2022vvn,Co:2020xaf,Arnold:2001ba,Arnold:2001ms}. In fact, it is sufficient to satisfy these conditions at $T_{\rm RH}$. Then $\phi$ decay will proceed as described above for $T \lesssim T_{\rm RH}$, which is the relevant range for multiple PTs in this scenario. We have checked that these conditions are satisfied for the blue dots shown in Fig.~\ref{fig:m-h}.

\begin{figure}[htbp!]
    \centering
    \includegraphics[width=0.55\textwidth]{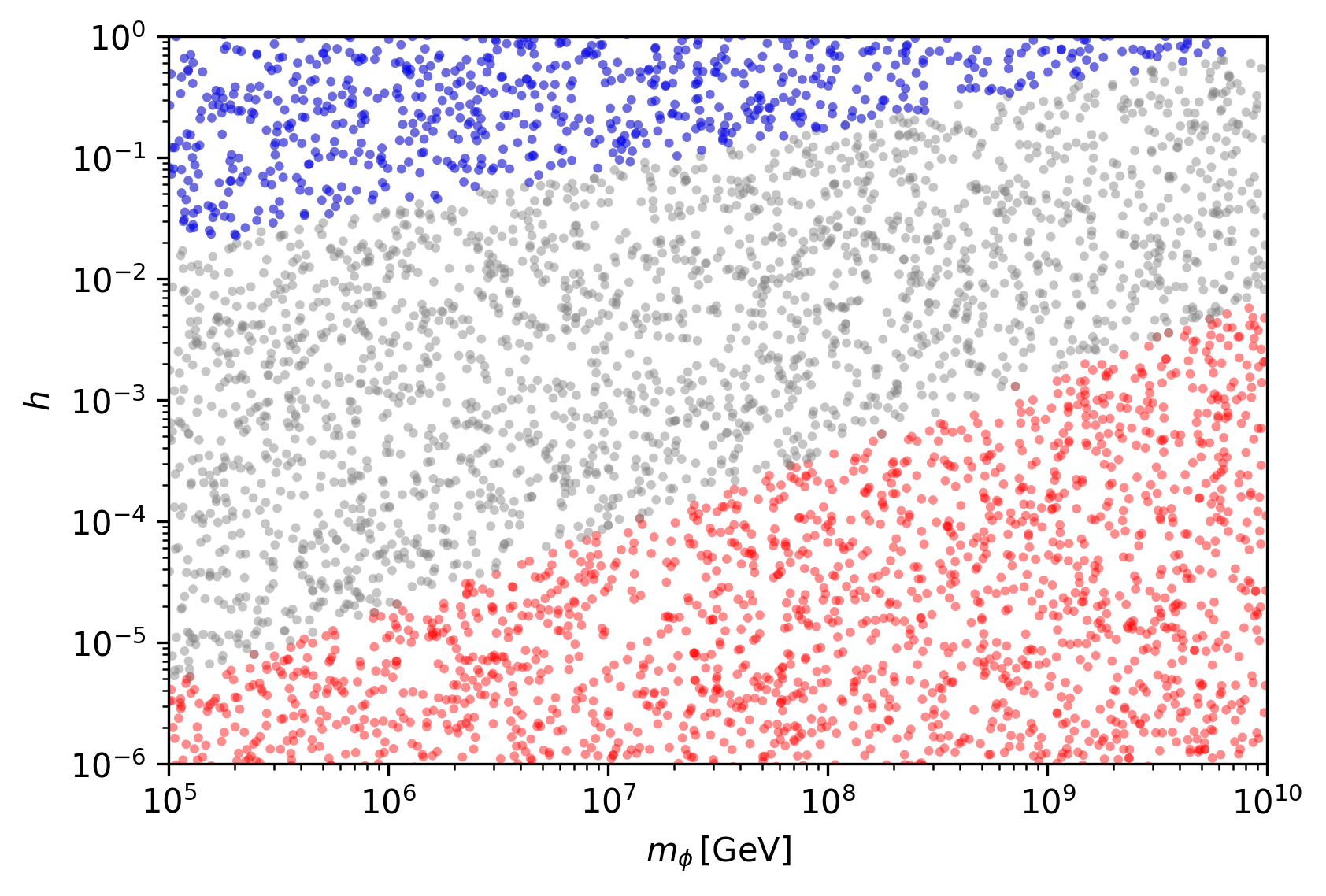}
    \caption{This figure depicts regions with different temperature evolution during the non-adiabatic phase of EMD in the $m_\phi-h$ plane. The dots are obtained from scans over $m_\phi$ and $h$ for the initial condition $T_{\rm O} = 10^{11}$ GeV. The blue dots correspond to cases where $T \propto a^{3/8}$ in the non-adiabatic phase followed by a smooth transition to $T \propto a^{-1}$ in RD. The gray dots represent cases where the same behavior is accompanied by a sharp increase in $T$ at the end of EMD. The red dots correspond to the standard case where $T \propto a^{-3/8}$ throughout the non-adiabatic phase of EMD. 
    }
    \label{fig:m-h}
\end{figure}

\section{Gravitational Waves from Multiple First-Order Phase Transitions}
\label{sec:pt-gw}

\subsection{General Formalism}

Cosmological PTs happen when symmetries are spontaneously broken as the early universe cools down. FOPTs typically lead to production of GW when bubbles of the new stable phase nucleate and percolate  
provided that their associated anisotropic stresses act coherently for (at least) the characteristic duration of 
 PT, denoted by $\beta^{-1}$, where $\beta$ is 
the slope of the bounce action along the thermal trajectory~\cite{Mazumdar:2018dfl,Weir:2017wfa,Athron:2023xlk}
\begin{eqnarray}
\label{eq:beta_def}
\beta \equiv
\left.
\frac{d}{dt}\ln\frac{\Gamma}{\mathcal V}
\right|_{T_{\rm PT}}
.
\end{eqnarray}
Here $\Gamma/\mathcal V$ denotes the thermal bubble nucleation rate per unit
physical volume, where $\mathcal V$ is a spatial volume.  A
general criterion for efficient nucleation is that the expected number of
bubbles nucleated in one Hubble volume during one Hubble time becomes
order unity~\cite{Mazumdar:2018dfl,Weir:2017wfa,Athron:2023xlk}.

GW are produced from three main sources during FOPTs: 
bubble collisions, sound waves, and turbulence in the primordial plasma
during the PT. Therefore, the GW spectrum at the time of production follows 
\begin{eqnarray}
\label{eq:tot_gw}
{\tilde \Omega_{\rm GW,*}}(f_*)
= {\tilde \Omega_{\mathrm{bc},*}} ~ {\cal S}_{\rm bc}(f_*)
+ {\tilde \Omega_{\mathrm{sw},*}} ~ {\cal S}_{\rm sw}(f_*)
+ {\tilde \Omega_{\mathrm{turb},*}} ~ {\cal S}_{\rm turb}(f_*)
\, ,
\end{eqnarray}
where the subscripts denote the amplitude of contributions from bubble collisions, sound waves in the plasma, and magnetohydrodynamic turbulence, respectively. The subscript $*$ denotes the value of a given quantity at the time of PT. The amplitude ${\tilde \Omega}_{\rm GW,*}$
shows how strongly a given component of the GW spectrum depends on the PT parameters, and the spectrum function $\mathcal{S}(f_*)$ specifies the frequency dependence of that component. Henceforth, we do not consider the contribution of turbulent GW spectrum since it is subdominant in cases of interest to us \cite{Mazumdar:2018dfl,Weir:2017wfa}. 

The existing analytical fit functions for GW from FOPTs are provided over a wide range of PT parameters \cite{Guo:2024gmu}. They can be used for theoretical predictions of different beyond the SM scenarios and the prospects for probing a wide range of viable PT parameters in these models by GW observatories such as LISA and DECIGO.

For the bubble wall collisions, one can use the envelope approximation~\cite{Kosowsky:1992rz, Kosowsky:1992vn, Huber:2008hg, Weir:2016tov}. The spectral shape and the normalized amplitude is given by~\cite{Caprini:2018mtu, Caprini:2019egz}
\begin{align}
\label{eq:bc_spec_reform}
 & \mathcal{S}_{\rm bc}(f_*) = \frac{3.8\, (f_*/f_{\rm bc,*})^{2.8}}{1 + 2.8\, (f_*/f_{\rm bc,*})^{3.8}}\,, \\    
& \widetilde{\Omega}_{\rm bc,*} = 
    \left(\frac{0.11\, v_w^3}{0.42 + v_w^2}\right) 
    \left(\frac{\kappa_{\phi}\, \alpha_{\rm *}}{1 + \alpha_{\rm *} + R_{\phi,*}}\right)^2 
    \left(\frac{\beta}{H_*}\right)^{-2}\,.
\end{align}
Here $v_w$ is velocity of the bubble wall and $\alpha_{\rm *}$ is the ratio of vacuum energy injected to the plasma that is an indicator of the transition strength.
The parameter $\kappa_{\phi}$ quantifies the efficiency of transferring vacuum energy to the kinetic energy of the bubble wall, which can be estimated as~\cite{Caprini:2018mtu, Caprini:2019egz}
\begin{equation}
\label{eq:kappaphi_reform}
    \kappa_{\phi} \approx \frac{0.18\, \sqrt{\alpha_{\rm *}} + 0.72\, \alpha_{\rm *}}{1 + 0.72\, \alpha_{\rm *}}\,.
\end{equation}

The GW background from sound waves arises from long-lived oscillations in a relativistic plasma. Hydrodynamic simulations with general equations of state set the spectral shape and normalization \cite{Hindmarsh:2015qta, Hindmarsh:2017gnf, Weir:2017wfa, Caprini:2018mtu, Caprini:2019egz}. 
In the  limit of bag model this will lead to a simple fit \cite{Steinhardt:1981ct, Espinosa:2010hh}. The spectrum is then given by~\footnote{We will use a more precise expression for the sound waves contribution to the spectrum as described in the next section.}
\begin{eqnarray} 
\label{sw}
&& \mathcal{S}_{\rm sw}(f_*) = \left(\frac{f_*}{f_{\rm sw,*}}\right)^3 \left(\frac{7}{4 + 3 (f_*/f_{\rm sw,*})^2}\right)^{7/2}, \\
&& \widetilde{\Omega}_{\rm sw, *} \approx 7.244 \times 10^{-2}\, v_w\,
\left(\frac{\kappa_{\rm sw}\, \alpha_{\rm *}}{1 + \alpha_{\rm *} + R_{\phi,*}}\right)^2\,
\left(\frac{\beta}{H_{*}}\right)^{-1}\,
\Upsilon(\omega) \, ,
\end{eqnarray}
where the parameter $\kappa_{\rm sw}$ is the efficiency in converting the released vacuum energy to bulk motion
\begin{equation}
\label{eq:kappa_sw}
 \kappa_{\rm sw} \approx \frac{\alpha_{\rm *}}{0.73 + 0.083\,\sqrt{\alpha_{\rm *}} + \alpha_{\rm *}} .   
\end{equation}
The suppression factor $\Upsilon(\omega)$ takes the finite duration of the acoustic waves $\tau_{\rm sw}$ into account~\cite{Ellis:2020awk, Guo:2020grp}. It shows how the effective lifetime of the sound waves is modified by the background equation of state~\cite{Guo:2024kfk, Hindmarsh:2019phv}. For MD ($\omega=0$) and RD ($\omega=\tfrac{1}{3}$) cases, one can show that~\cite{Banik:2025olw,Xiao:2024rsj}
\begin{eqnarray}
\label{eq:ups_0}
\Upsilon_{\omega=0}
&=& \frac{2}{3}\left[1 - \left(1+\frac{3}{2}\,\tau_{\rm sw}H_*\right)^{-1}\right] \,,
\end{eqnarray}
\begin{eqnarray}
\tau_{\rm sw}\,H_* \big|_{\omega=0}&\simeq& \frac{(8\pi)^{1/3}\,v_w}{\bar{U}_f}\left(\frac{\beta}{H_*}\right)^{-1} \,,
\end{eqnarray}
\begin{eqnarray}
\bar{U}_f^{\,2}\big|_{\omega=0}
&=& \frac{3}{4}\left[\frac{\kappa_{\rm sw}\alpha_*(1+R_{\phi,*})}{1+\alpha_*+R_{\phi,*}}\right]
\left(1+\frac{R_{\phi,*}}{4}\right)^{-1} .
\end{eqnarray}
and
\begin{eqnarray}
\label{eq:ups_13}
\Upsilon_{\omega=1/3}
&=& 1 - \left(1+2\,\tau_{\rm sw}H_*\right)^{-1/2} \,,
\end{eqnarray}

\begin{eqnarray}
\tau_{\rm sw}\,H_* \bigg|_{\omega=1/3}&\simeq& \frac{(8\pi)^{1/3}\,v_w}{\bar{U}_f}\left(\frac{\beta}{H_*}\right)^{-1} \,,
\end{eqnarray}

\begin{eqnarray}
\bar{U}_f^{\,2}\bigg|_{\omega=1/3}
&=& \frac{3}{4}\left[\frac{\kappa_{\rm sw}\alpha_*(1+R_{\phi,*})}{1+\alpha_*+R_{\phi,*}}\right]
\left(1+\frac{R_{\phi,*}}{3}\right)^{-1} .
\end{eqnarray}
In above equations the ratio 
scalar field to radiation density at phase transition temperature  is given by  

\begin{eqnarray}
\label{eq:ratio}
R_{\phi,*} &\equiv& \frac{\rho_\phi}{\rho_{\rm R}} \bigg|_{T_{\rm PT}} .
\end{eqnarray}
Faster transitions with larger $\beta/H_{*}$ have a shorter bubble lifetime leading to a stronger suppression of the GW spectrum~\footnote{ 
In Ref.~\cite{Ellis:2020awk,Hindmarsh:2017gnf}, the source lifetime is approximated by the characteristic fluid length scale (comparable to the mean bubble separation) divided by the relevant velocity~\cite{Hindmarsh:2017gnf}. The general expressions like above have been studied for arbitrary equation of state parameters~\cite{Guo:2024kfk, Hindmarsh:2019phv}.
}.

Peak frequencies for bubble collision and sound waves and contributions can be estimated using the details of PT ~\cite{Caprini:2018mtu, Athron:2023xlk}
\begin{align}
    f_{\rm bc,*} &\approx \frac{0.62\,\beta}{1.8 - 0.1 v_w + v_w^2}\,, \\[6pt]
    f_{\rm sw,*} &\approx \frac{2\,\beta}{\sqrt{3}\,v_w}\,.
    \label{eq:fstar_reform}
\end{align}
In scenarios where the system evolves in the so-called detonation regime~\cite{Espinosa:2010hh, Ellis:2018mja, Hindmarsh:2019phv}, the wall velocity can be reliably approximated in the following analytic form 
\begin{equation}
\label{eq:wall_vel_reform}
    v_w \approx 
    \frac{ \tfrac{1}{\sqrt{3}} + \sqrt{ \alpha_{\rm *}^2 + \tfrac{2}{3}\alpha_{\rm *} } }
         { 1 + \alpha_{\rm *} } \,.
\end{equation}
However, in order to explore the more general situations that are possible in our scenario, we may consider $v_w$ as a free parameter.

\begin{figure}
    \centering
\begin{tabular}{ccc}
\includegraphics[width=0.28\textwidth]{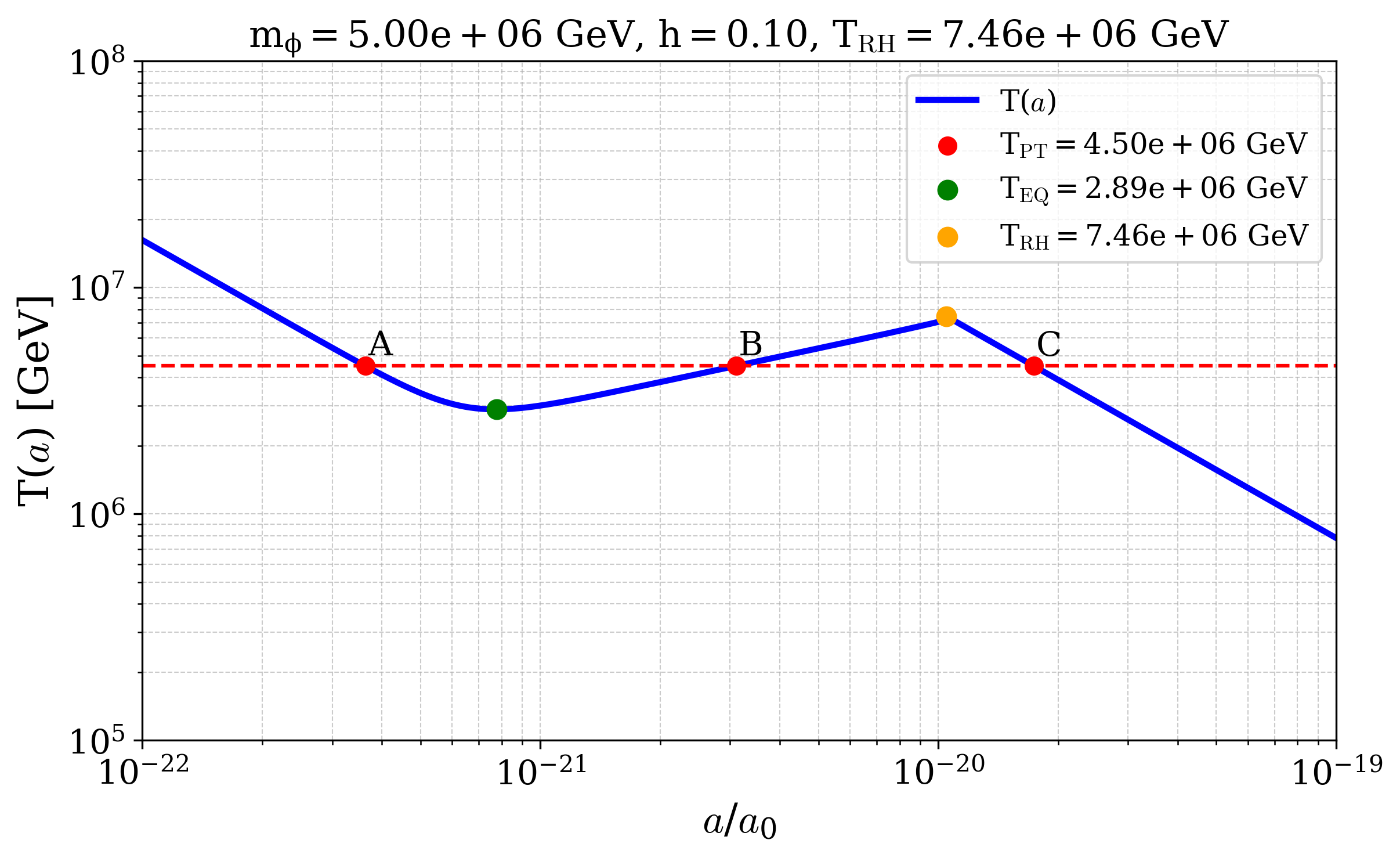}
&
\includegraphics[width=0.28\textwidth]{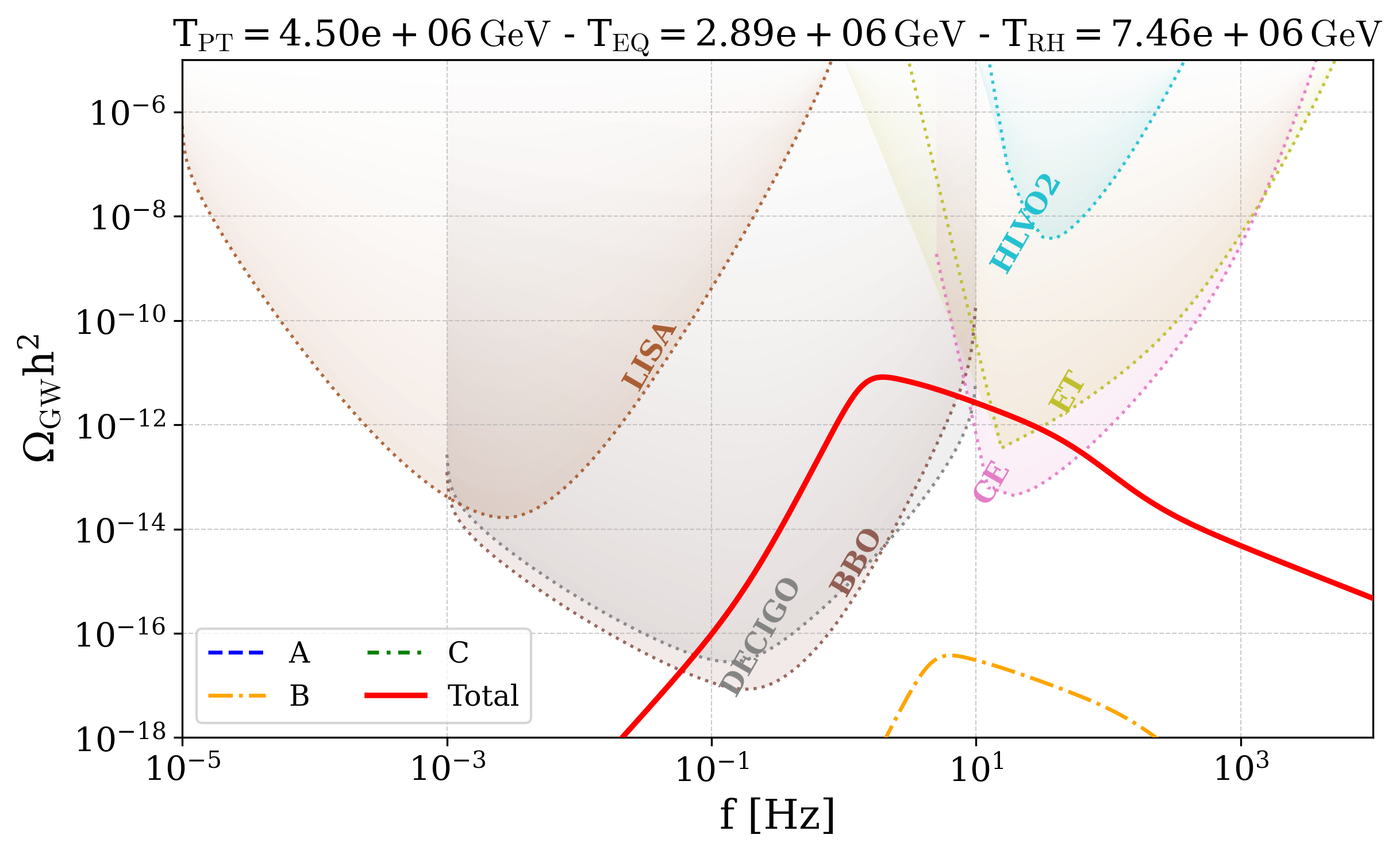}
&
\includegraphics[width=0.28\textwidth]{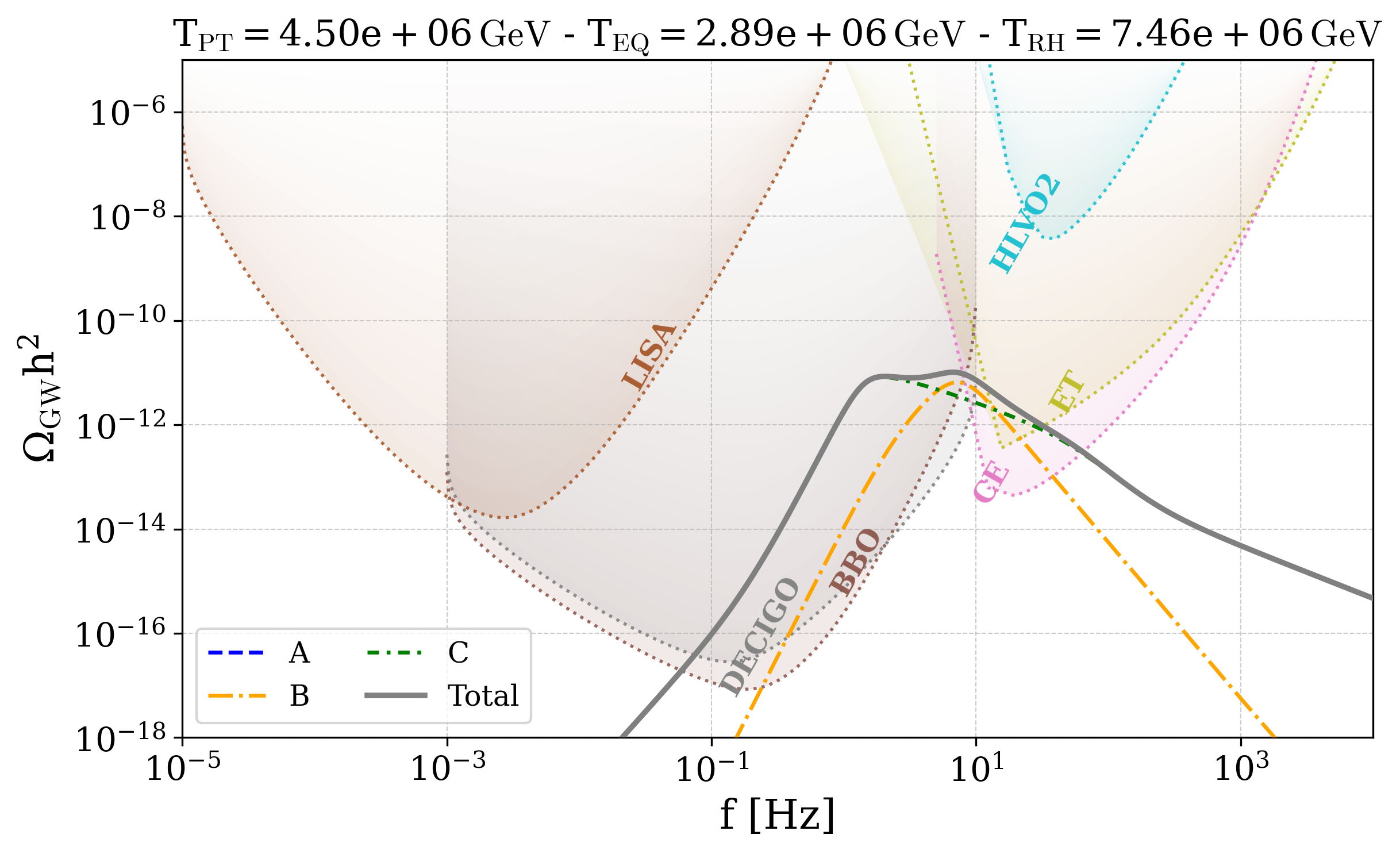}
\\[1.0em]

\includegraphics[width=0.28\textwidth]{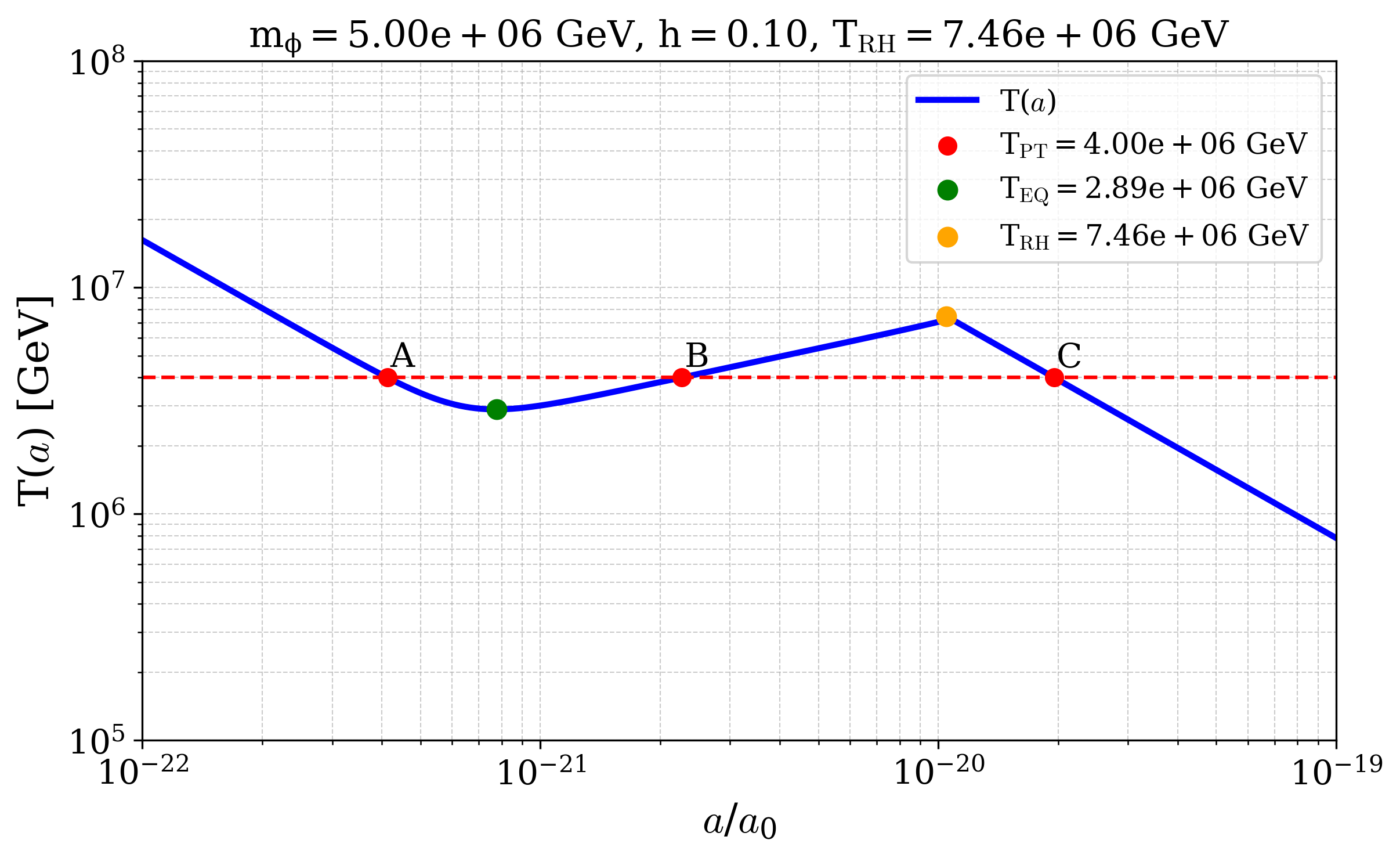}
&
\includegraphics[width=0.28\textwidth]{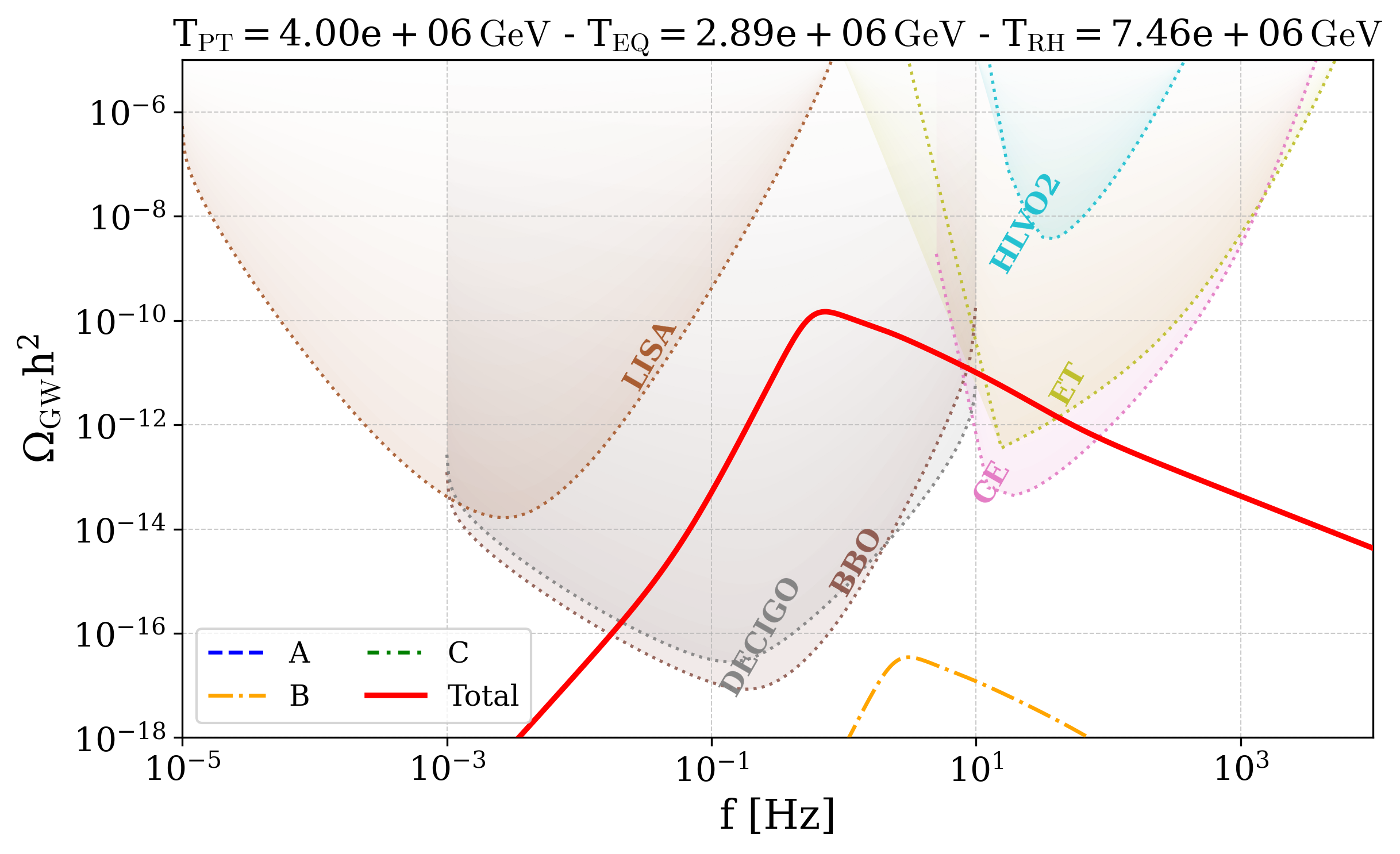}
&
\includegraphics[width=0.28\textwidth]{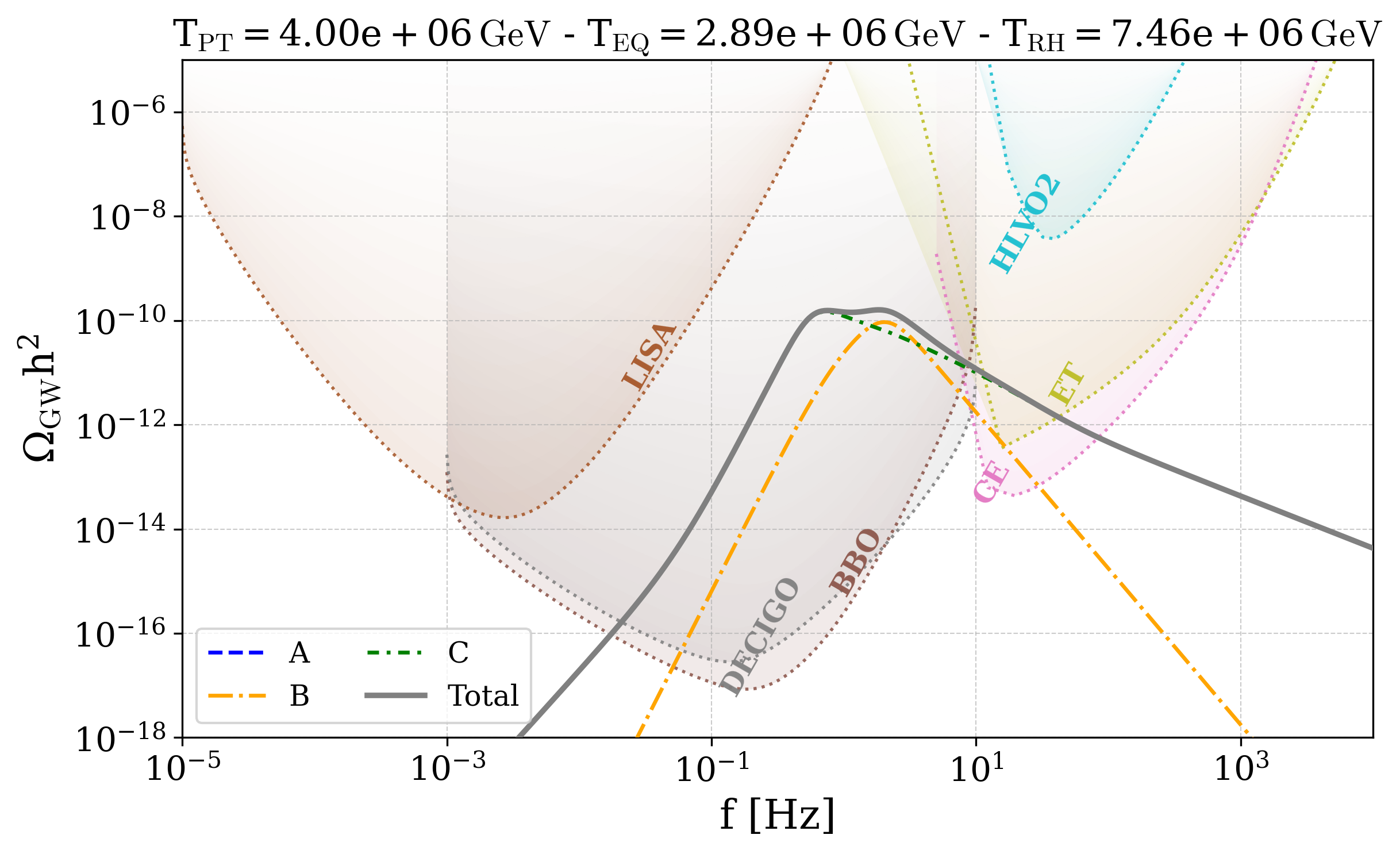}
\\[1.0em]

\includegraphics[width=0.28\textwidth]{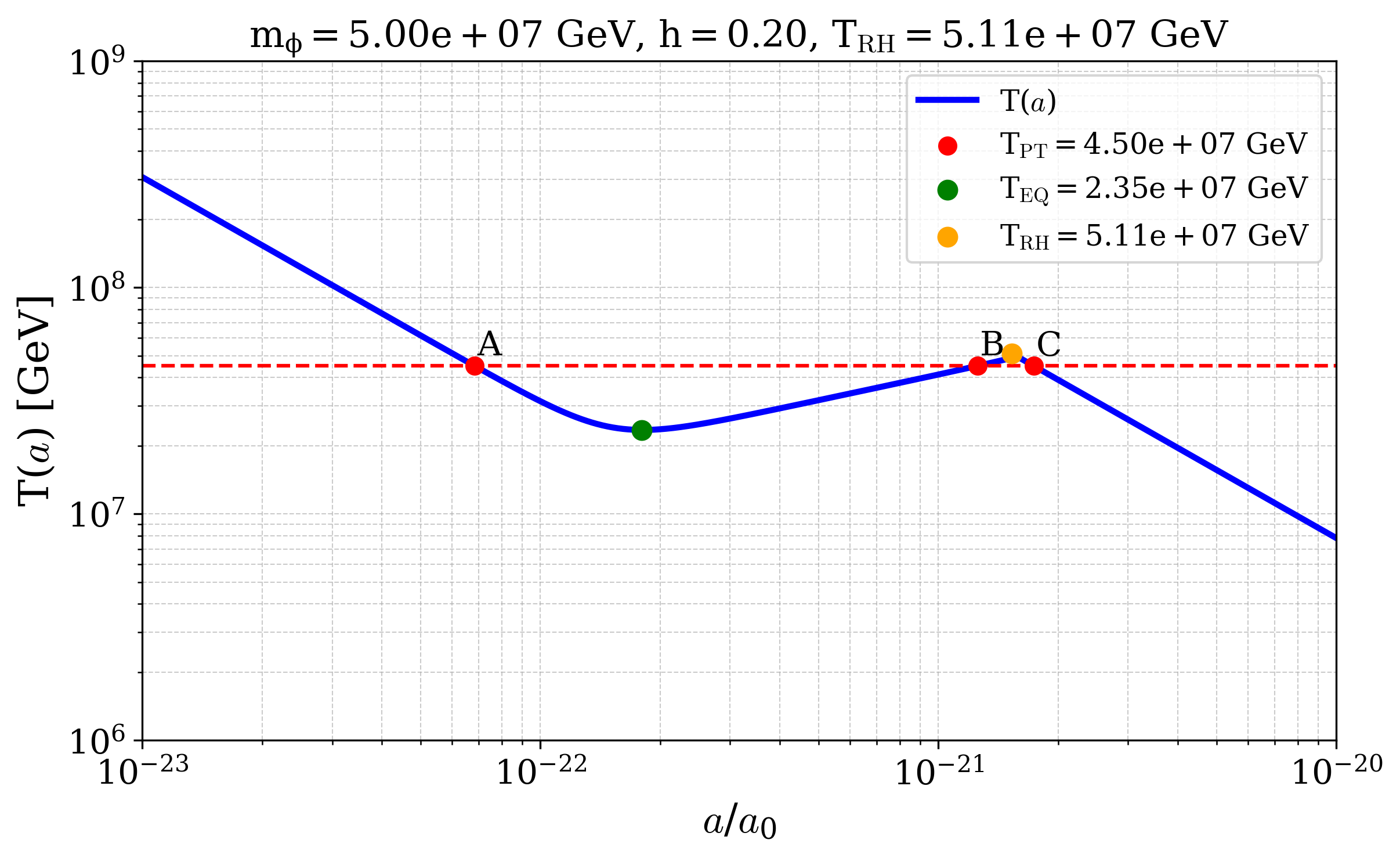}
&
\includegraphics[width=0.28\textwidth]{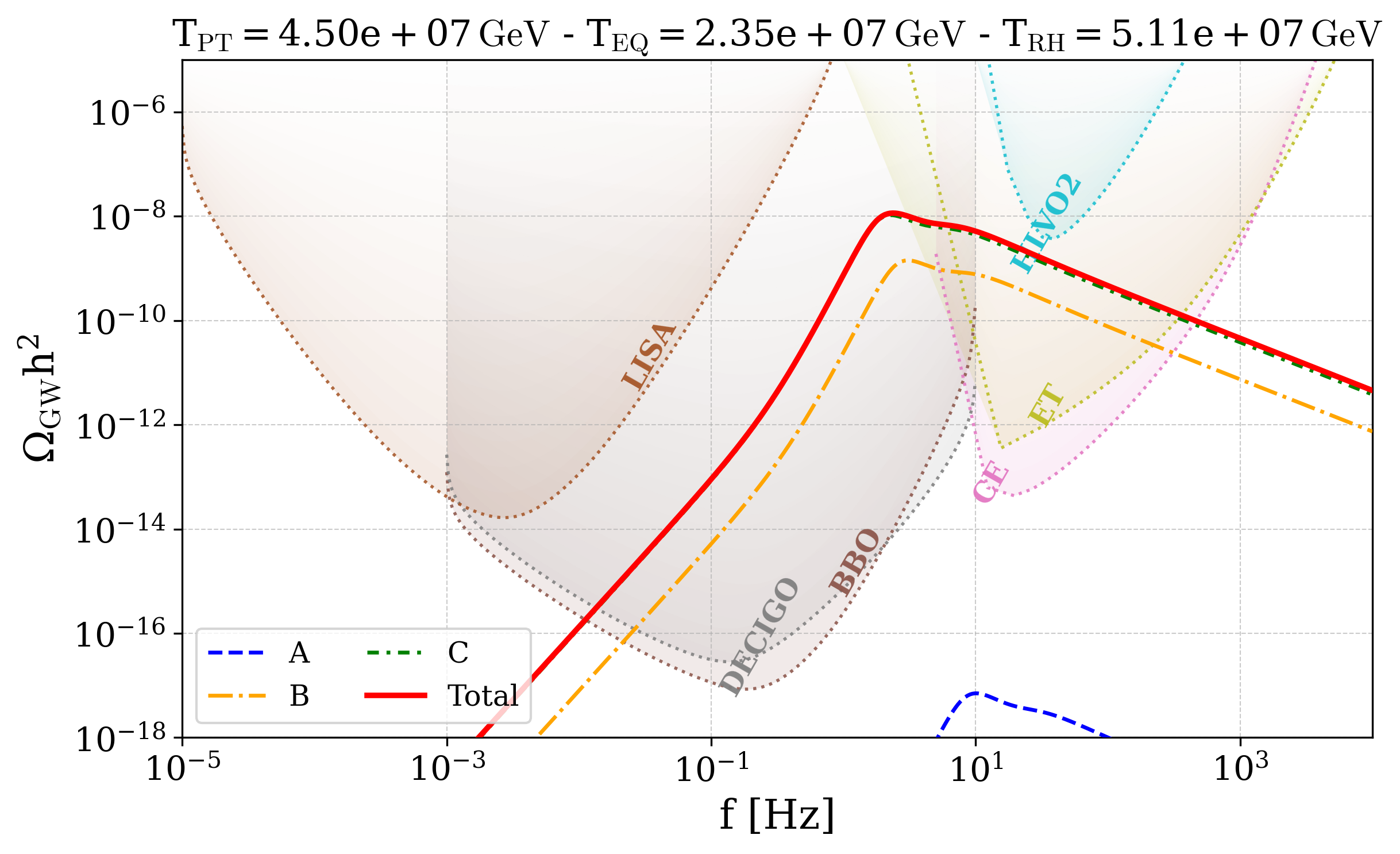}
&
\includegraphics[width=0.28\textwidth]{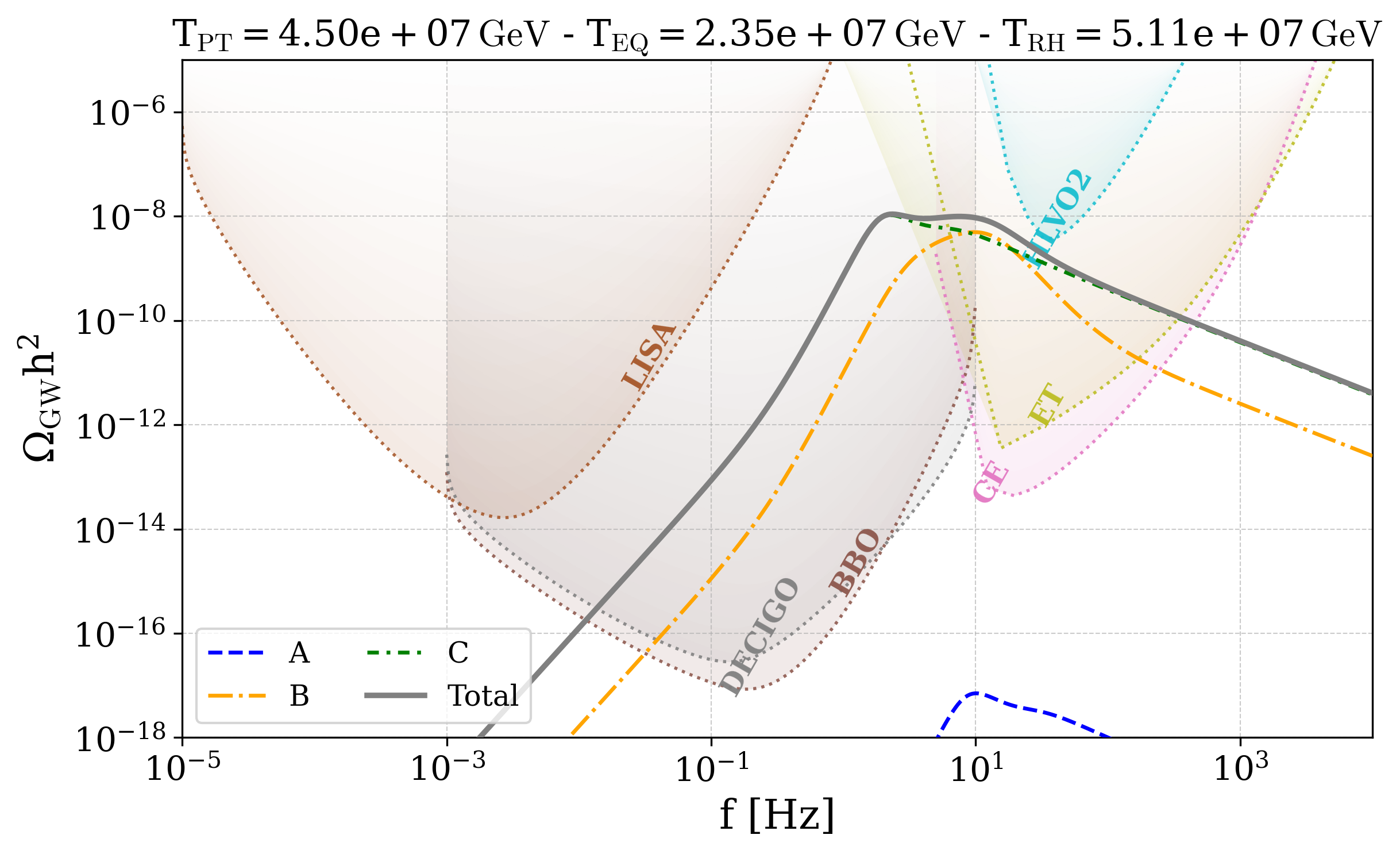}
\\[1.0em]

\includegraphics[width=0.28\textwidth]{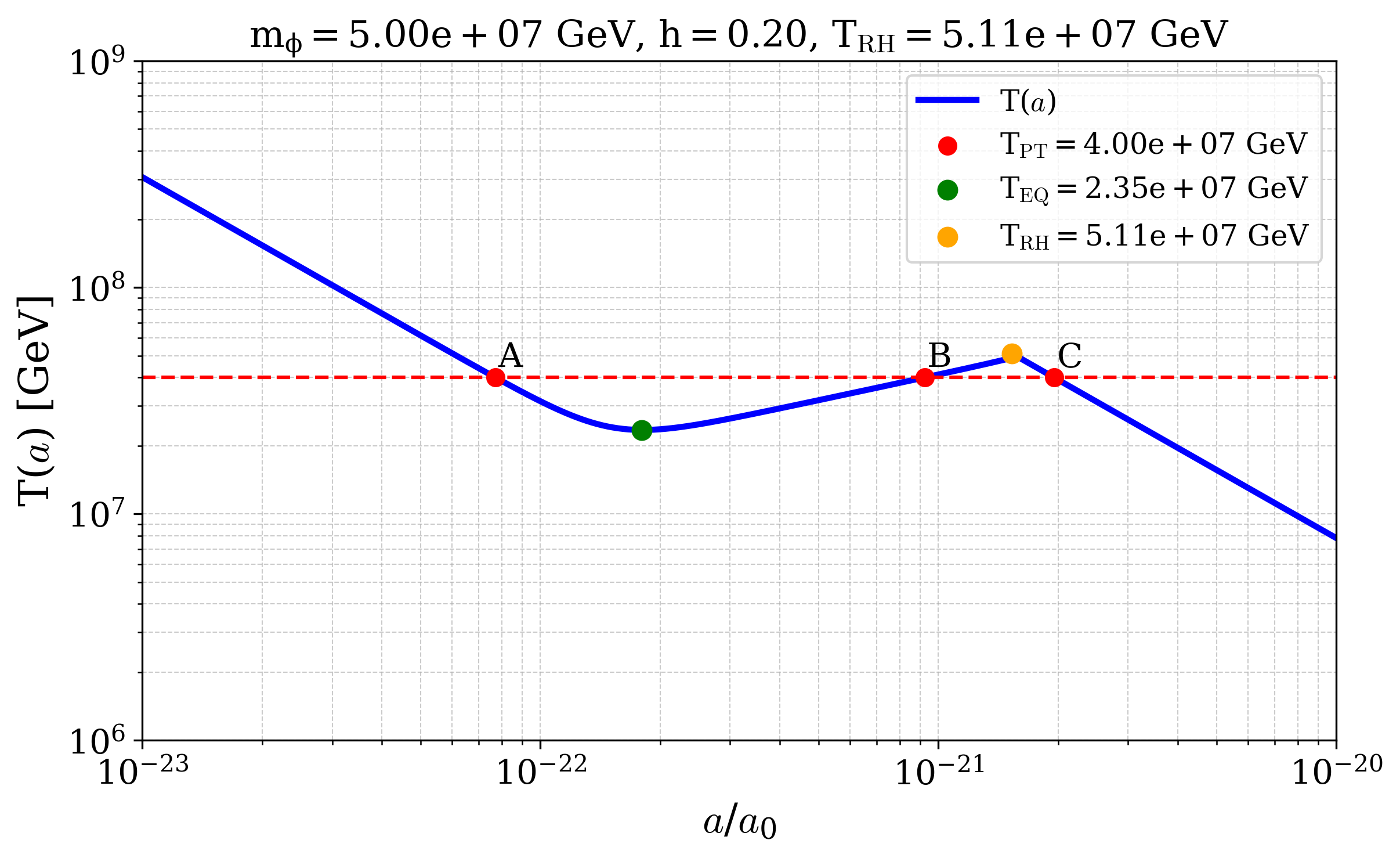}
&
\includegraphics[width=0.28\textwidth]{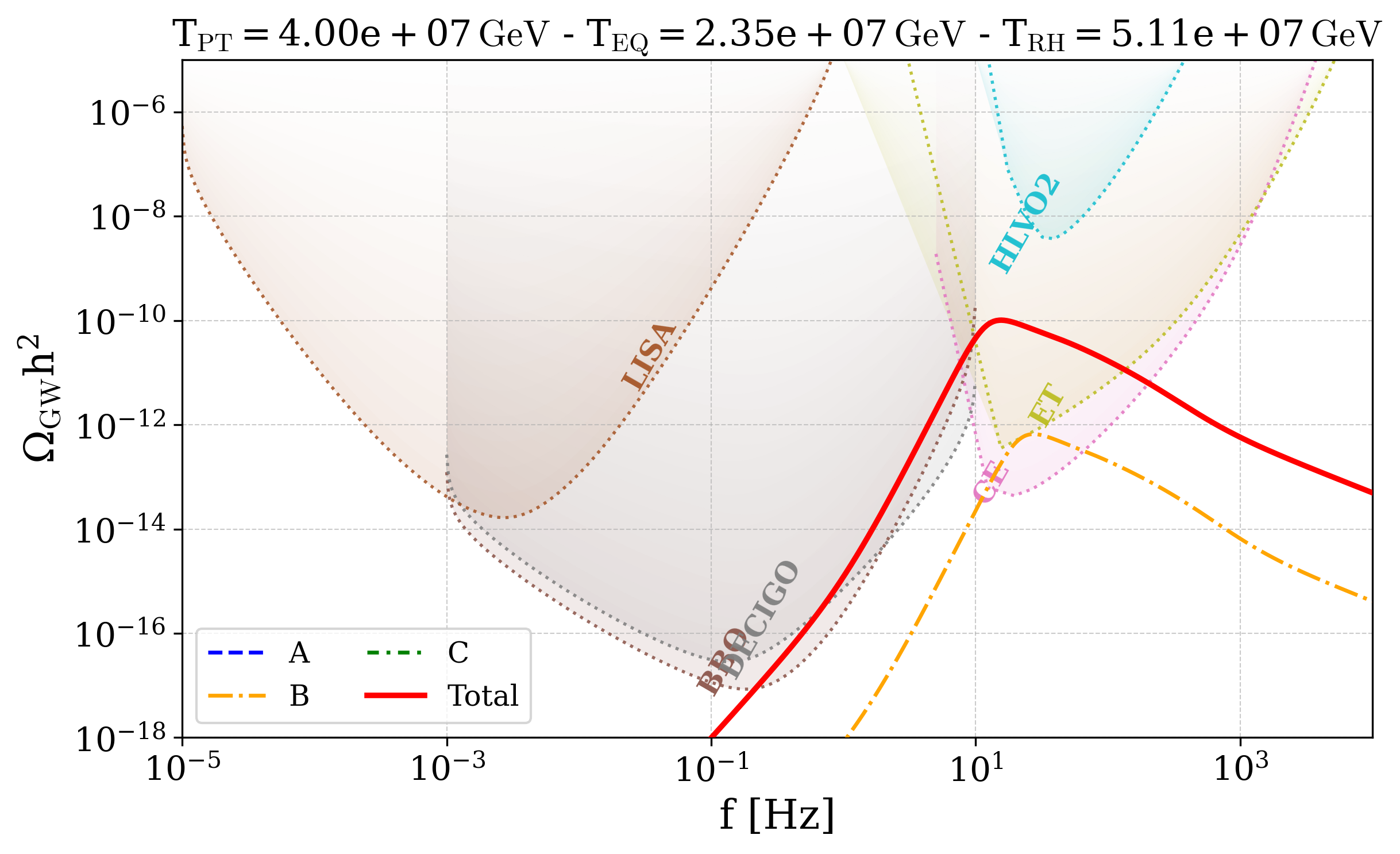}
&
\includegraphics[width=0.28\textwidth]{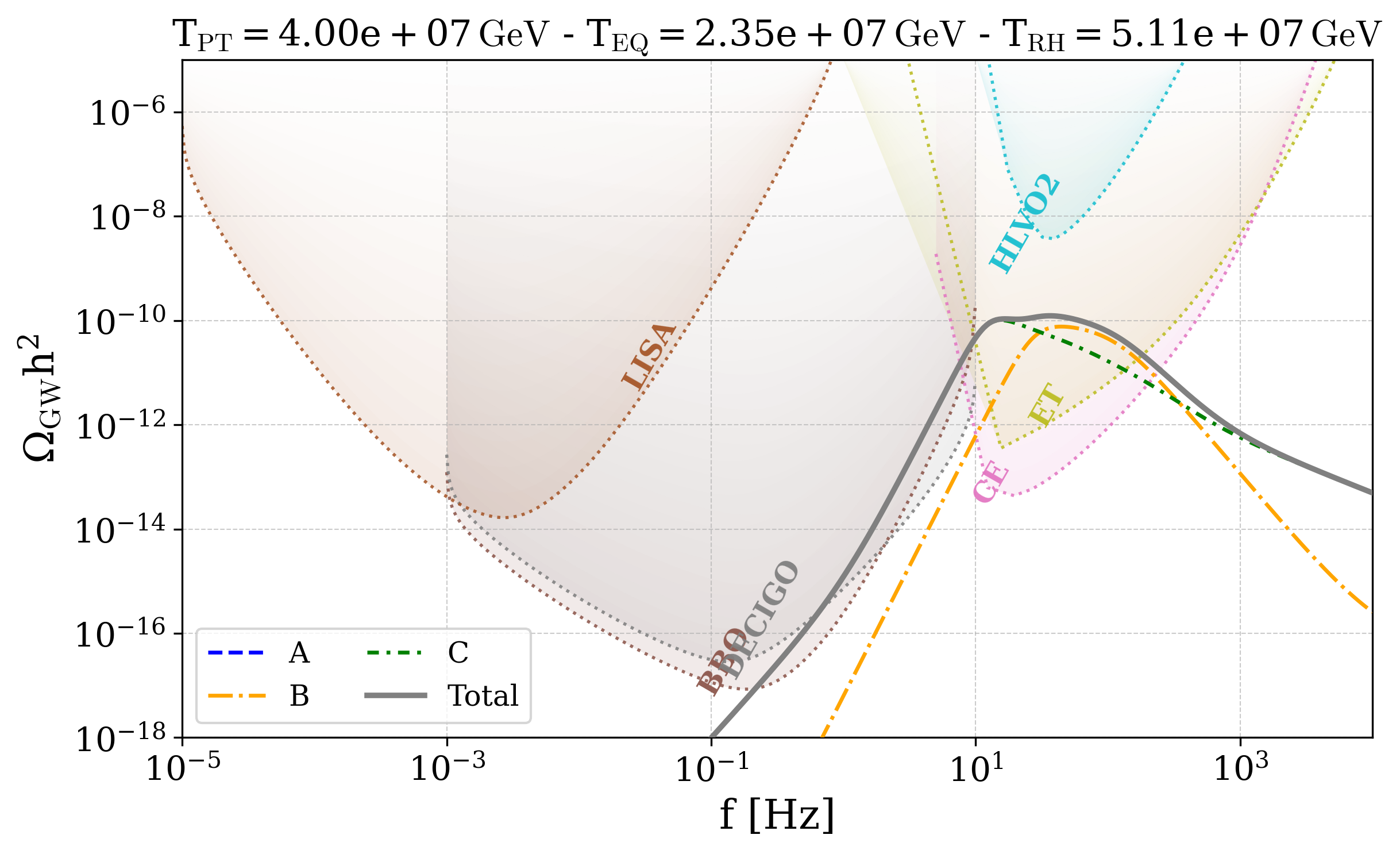}
\\[1.0em]

\includegraphics[width=0.28\textwidth]{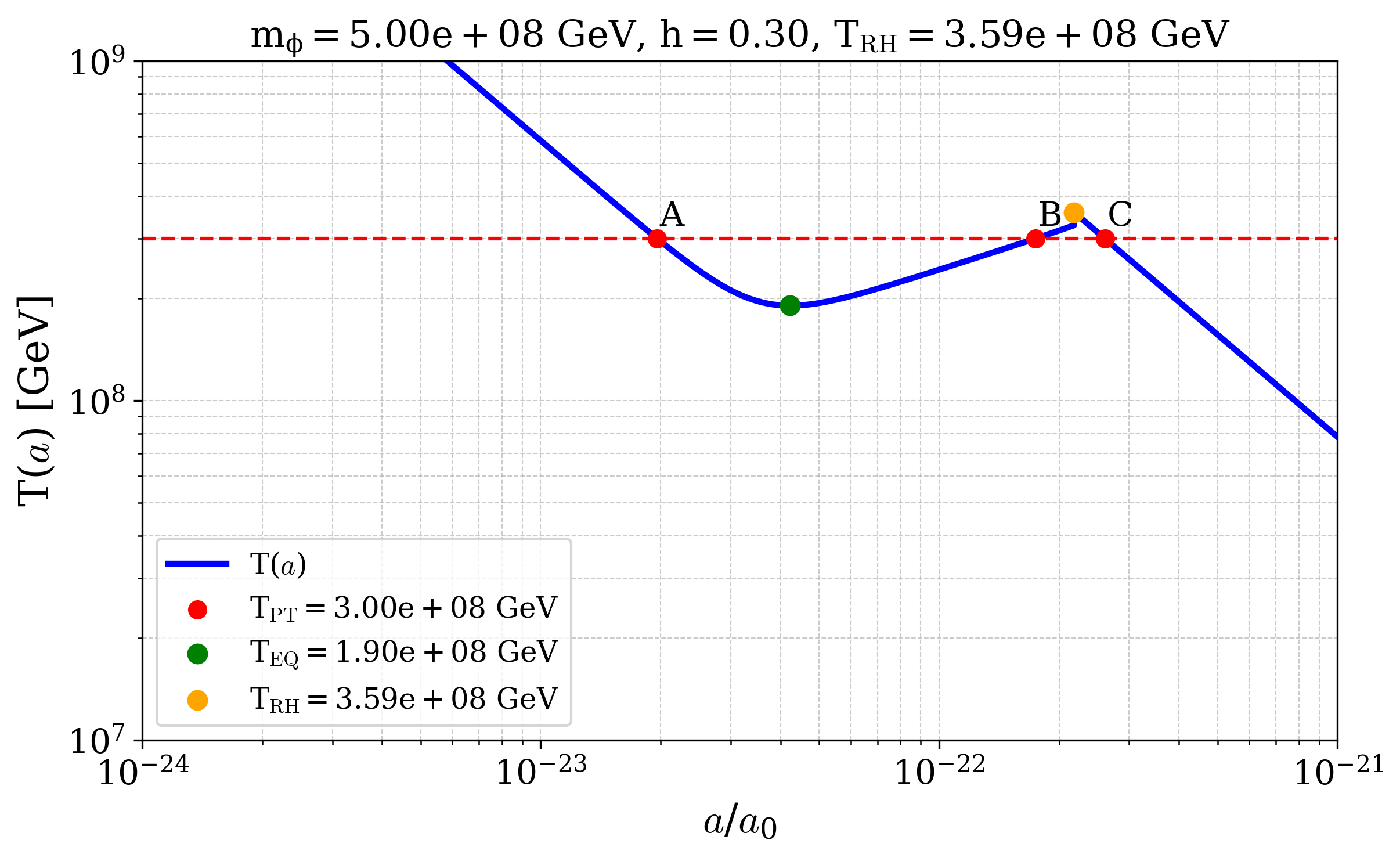}
&
\includegraphics[width=0.28\textwidth]{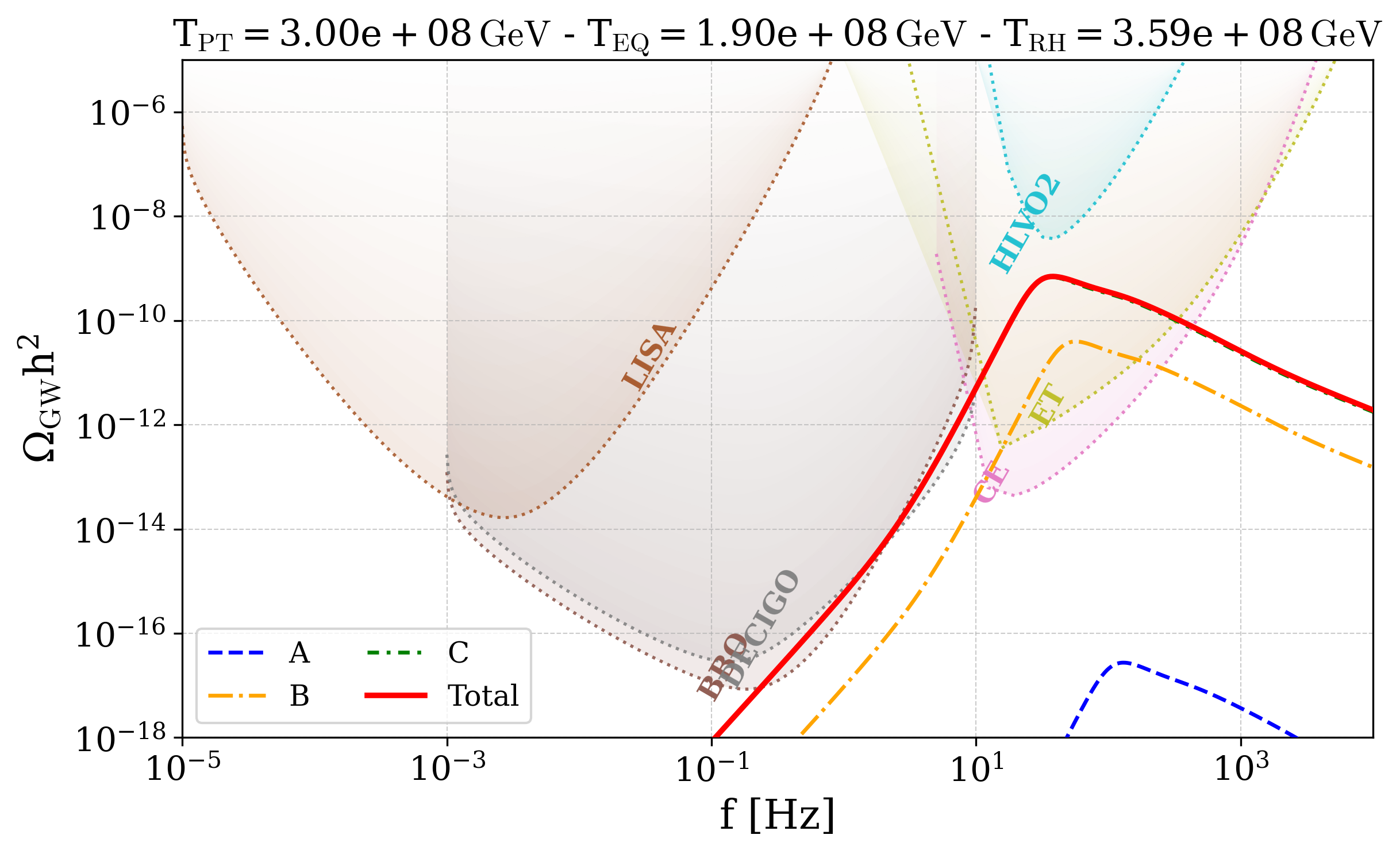}
&
\includegraphics[width=0.28\textwidth]{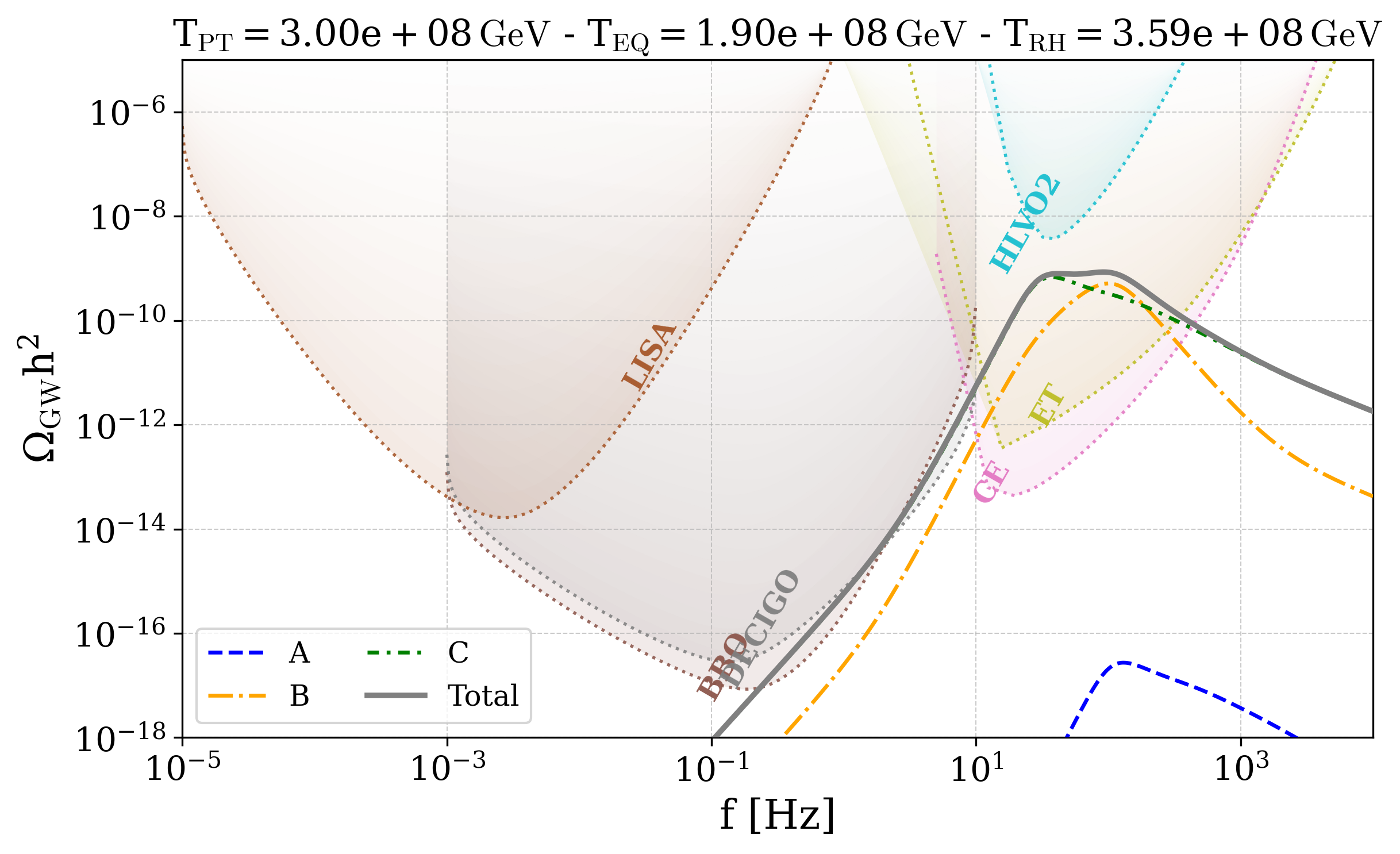}
\\[1.0em]

\includegraphics[width=0.28\textwidth]{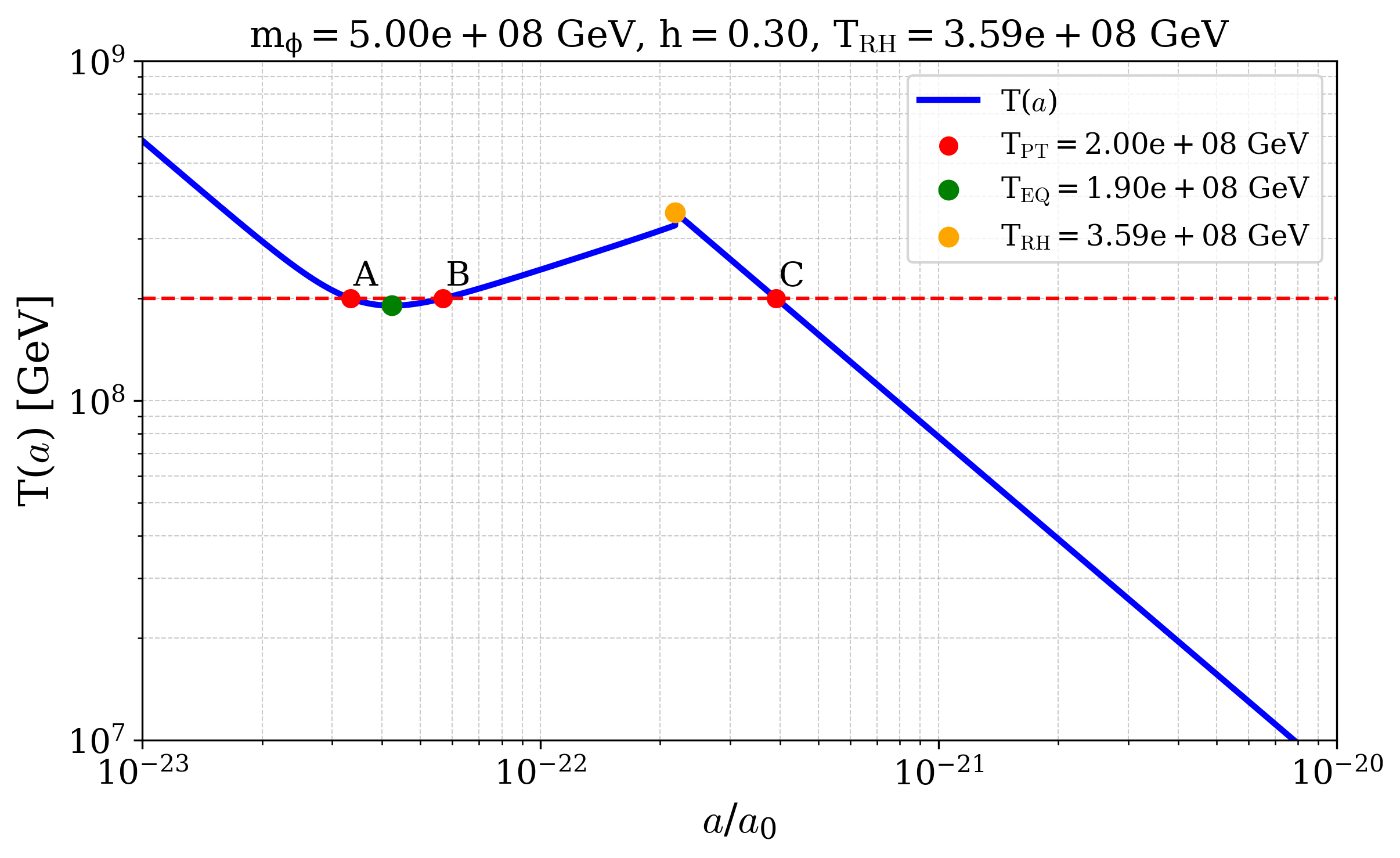}
&
\includegraphics[width=0.28\textwidth]{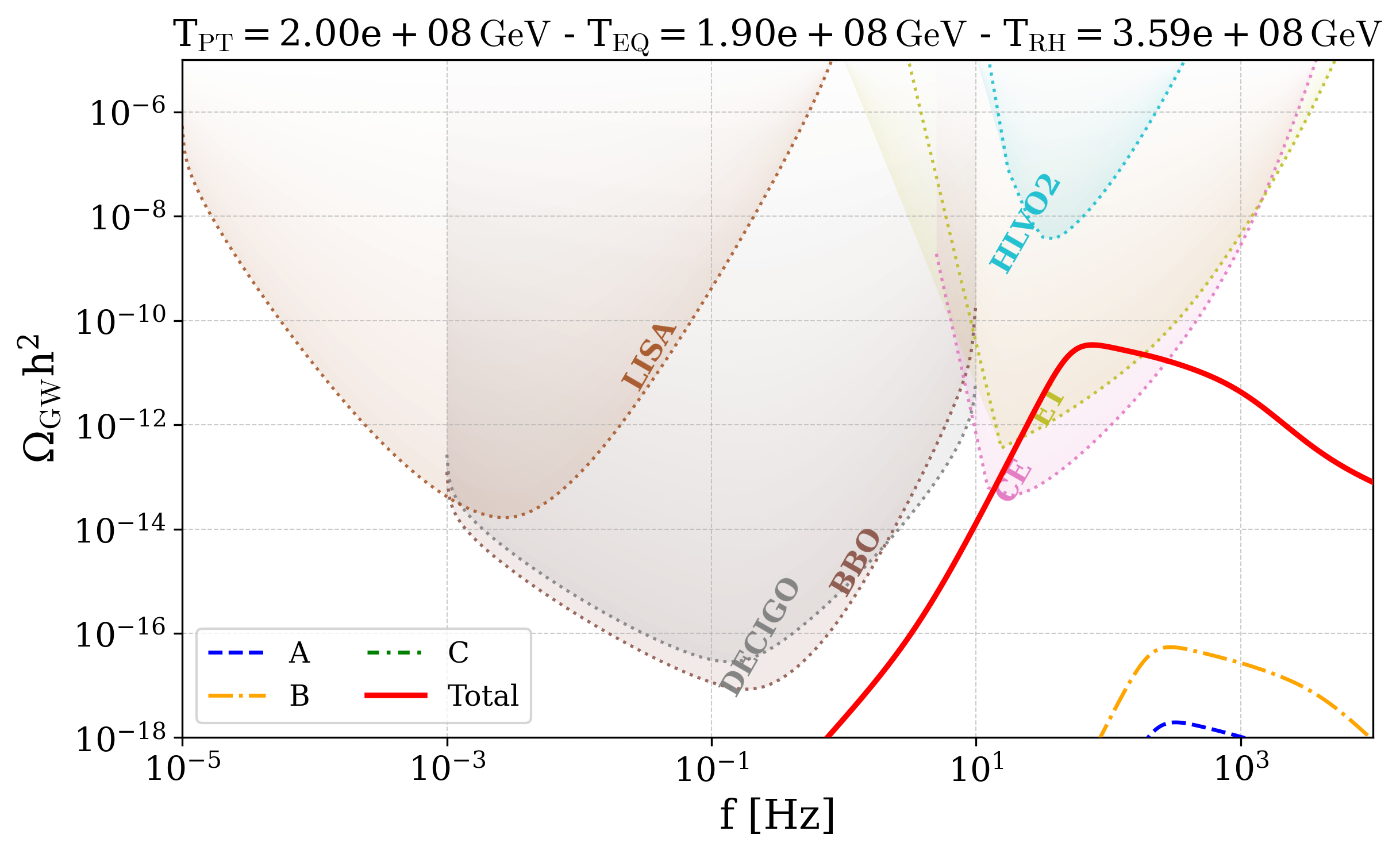}
&
\includegraphics[width=0.28\textwidth]{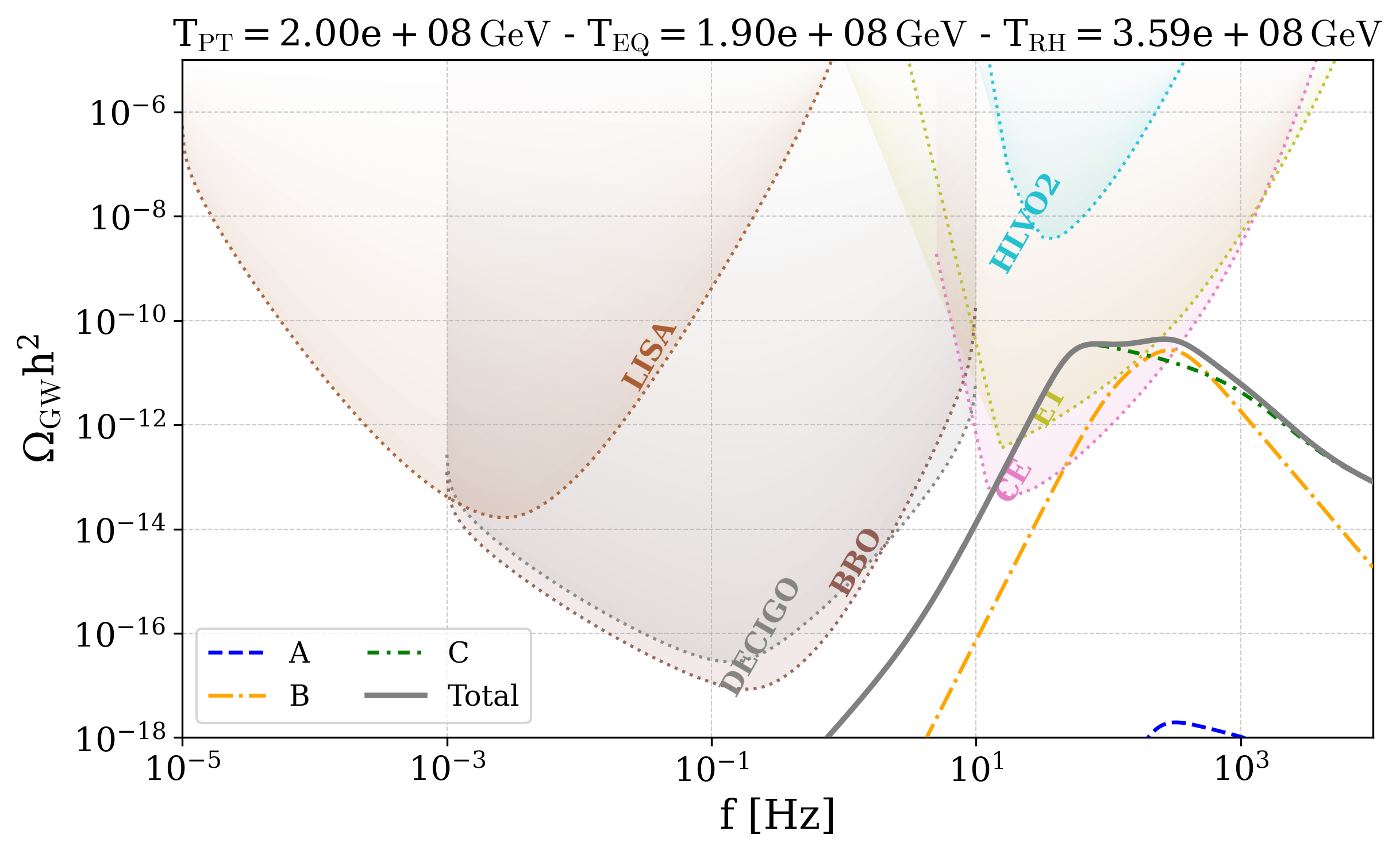}
\end{tabular}
    \caption{The results from multiple FOPTs for 6 benchmark points given in Table~\ref{tab:params}. Left panels show temperature evolution as a function of the scale factor for these points. Middle panels show the individual contributions from each transition (dashed) and the total GW spectrum (solid red) for the standard choice of PT parameters. The individual contributions (dashed) and the total spectrum (solid gray) for the alternative choice of PT parameters are shown in the right panels.  
    The experimental sensitivity curves in the GW spectra plots are taken from the data of Ref.~\cite{Schmitz:2020syl}.
    }
    \label{fig:hphi-temp-gw-tot}
\end{figure}

\begin{table}[t]
\centering
\begin{minipage}{0.98\textwidth}
\centering
\setlength{\tabcolsep}{2.4pt}
\renewcommand{\arraystretch}{1.12}
\resizebox{\linewidth}{!}{
\begin{tabular}{c c c c c c c c}
\hline
$m_\phi\,[{\rm GeV}]$ &
$h$ &
$T_{\rm PT}\,[{\rm GeV}]$ &
$T_{\rm EQ}\,[{\rm GeV}]$ &
$T_{\rm RH}\,[{\rm GeV}]$ &
$\alpha_*$ &
$v_w$ &
$\beta/H_*$ \\
\hline
$5.0\times10^{6}$ & $0.10$ & $4.5\times10^{6}$ & $2.89\times10^{6}$ & $7.46\times10^{6}$ & $0.15\,(0.71)$ & $0.76\,(0.12)$ & $21.65\,(16.68)$ \\
$5.0\times10^{6}$ & $0.10$ & $4.0\times10^{6}$ & $2.89\times10^{6}$ & $7.46\times10^{6}$ & $0.27\,(0.81)$ & $0.74\,(0.06)$ & $11.40\,(4.67)$ \\
$5.0\times10^{7}$ & $0.20$ & $4.5\times10^{7}$ & $2.35\times10^{7}$ & $5.11\times10^{7}$ & $0.93\,(0.85)$ & $0.87\,(0.49)$ & $4.00\,(3.68)$ \\
$5.0\times10^{7}$ & $0.20$ & $4.0\times10^{7}$ & $2.35\times10^{7}$ & $5.11\times10^{7}$ & $0.37\,(0.77)$ & $0.77\,(0.19)$ & $26.24\,(29.00)$ \\
$5.0\times10^{8}$ & $0.30$ & $3.0\times10^{8}$ & $1.90\times10^{8}$ & $3.59\times10^{8}$ & $0.46\,(0.37)$ & $0.75\,(0.37)$ & $9.30\,(4.46)$ \\
$5.0\times10^{8}$ & $0.30$ & $2.0\times10^{8}$ & $1.90\times10^{8}$ & $3.59\times10^{8}$ & $0.25\,(0.75)$ & $0.69\,(0.06)$ & $24.34\,(13.63)$ \\
\hline
\end{tabular}
}
\end{minipage}
\caption{Parameter choices for the 6 benchmark points described in the text. 
The first two columns show parameters of the particle physics model described in Section~\ref{sec:bfeqs}. The next three columns show the transition temperature $T_{\rm PT}$ as well as EMD parameters $T_{\rm EQ}$ and $T_{\rm RH}$ for the chosen values of $m_\phi$ and $h$ and the initial condition $T_{\rm O}=10^{12}\,\mathrm{GeV}$. The last three columns depict the PT parameters; the numbers inside parentheses correspond to alternate choices for the heating transition $B$.}
\label{tab:params}
\end{table}

\subsection{Gravitational Waves Spectrum from Cooling and Heating Phase Transitions}

In our scenario, the temperature decreases from $T_{\rm O}$ to  $T_{\rm EQ}$ in the adiabatic phase of EMD, followed by a rise from $T_{\rm EQ}$ to $T_{\rm RH}$ in the non-adiabatic phase. As a result of this non-monotonic behavior, transitions in both cooling and heating phases can occur if $T_{\rm EQ} < T_{\rm PT} < T_{\rm RH}$. In fact, one will have three PTs in this case:
\vskip 1.5mm
\noindent
{\bf (1)} A cooling transition during the adiabatic phase of EMD ($A$).
\vskip 1.5mm
\noindent{\bf (2)} A heating transition in the non-adiabatic phase of EMD ($B$).
\vskip 1.5mm
\noindent{\bf (3)} A second cooling transition in the RD phase established at the end of EMD ($C$).
\vskip 1.5mm

As shown in the literature \cite{Buen-Abad:2023hex,Dent:2024bhi,Barni:2024lkj}, cooling transitions are frictional and produce acoustic waves that propagate in the thermal bath that are dominated by sound waves with $v_w \ll 1$ and $\alpha_{\rm *} \lesssim \mathcal{O}(1)$. 
Heating transitions are frictionless, leading to  $v_w\to 1$, and reduce the sound waves lifetime. These are typically dominated by bubble collisions provided that the transition completes without vacuum domination  \cite{Weir:2017wfa,Weir:2016tov}.

The relic density of GW, from a given source, at present can be written as \cite{Banik:2025olw}
\begin{eqnarray}
\Omega_{\rm GW}(f)  &=& \mathcal{D}\,\tilde{\Omega}_{\rm GW,*}\,\mathcal{S}(f_*) \, .
\end{eqnarray}
The factor $\mathcal{D}$ accounts for the change in the amplitude due to cosmic expansion between the time of PT and today, and $f$ is the redshifted value of $f_*$ at the present time. 

In general, we have (see Appendix \ref{app:peak-ratio})
\begin{equation}
f=\frac{a_{*}}{a_0}\,f_{*}  \qquad , \qquad
\mathcal{D} =
\left(\frac{a_{*}}{a_0}\right)^{4}\left(\frac{H_{*}}{H_0}\right)^{2} 
.
\label{eq:general_redshift_Omega_f_app}
\end{equation}
We see from Eq.~(\ref{eq:fstar_reform}) that, for a given value of $\beta/H_*$, the peak frequencies of the contributions from sound waves and bubble collisions (at the time of PT) are $\sim H_*$. This implies that
\begin{eqnarray}
&& f^{\rm peak}_A \sim {a_A \over a_0} H_A \,, \\
&& f^{\rm peak}_B \sim {a_B \over a_0} H_B \, ,\\
&& f^{\rm peak}_C \sim {a_C \over a_0} H_C \, . 
\end{eqnarray}
We note that the peak frequency from the cooling transition in the RD phase (C) experiences the same redshift as in the standard thermal history, resulting in  
\begin{equation}
f_C^{\rm peak}
\sim
\left(\frac{\pi^2}{90}\right)^{1/2} g^{1/6}_{*,C}
g^{1/3}_{*,0} ~
\frac{T_0 T_{\rm PT}}{M_{\rm P}} \,, 
\end{equation}
where numerically we have $f_C^{\rm peak} \approx 1.6\times 10^{-7}\,{\rm Hz}\,(T_{\rm PT}/{\rm GeV})\,(g_{*,C}/100)^{1/6}$ up to an $\mathcal{O}(\beta/H_*)$ factor.
By contrast, those from the other PTs (A and B) feel the non-standard expansion history since part of their evolution takes place in the EMD epoch. As shown in Appendix C, see Eqs.~(\ref{peak1},\ref{peak2}), we find
\begin{eqnarray} \label{peaks}
&& f^{\rm peak}_A \sim \left({T_{\rm PT} \over T_{\rm EQ}}\right)^{11/6} \left({T_{\rm RH} \over T_{\rm PT}}\right)^{7/3} f^{\rm peak}_C \, ,\\
&& f^{\rm peak}_B \sim \left({T_{\rm RH} \over T_{\rm PT}}\right)^{7/3} f^{\rm peak}_C \, ,
\end{eqnarray}
\noindent
which implies that $f^{\rm peak}_C < f^{\rm peak}_B < f^{\rm peak}_A$.

As for the suppression factor $\mathcal{D}$, we have
\begin{eqnarray}
&& \mathcal{D}_A = \left({a_A \over a_0}\right)^4 \left({H_A \over H_0}\right)^2  \,, \\
&& \mathcal{D}_B = \left({a_B \over a_0}\right)^4 \left({H_B \over H_0}\right)^2 \, ,\\
&& \mathcal{D}_C = \left({a_C \over a_0}\right)^4 \left({H_C \over H_0}\right)^2 \, .
\end{eqnarray}
As shown in Appendix C, see Eq.~(\ref{suppression1},\ref{suppression2}), this results in 
\begin{eqnarray} \label{amplitudes}
&& \mathcal{D}_A \sim \left({T_{\rm EQ} \over T_{\rm PT}}\right)^{11/3} \left({T_{\rm PT} \over T_{\rm RH}}\right)^{8/3} \mathcal{D}_C 
\, , \\
 && \mathcal{D}_B \sim \left({T_{\rm PT} \over T_{\rm RH}}\right)^{8/3} \mathcal{D}_C\, ,
\end{eqnarray}
where for the cooling transition in the RD phase (C), we have 
\begin{equation}
\mathcal{D}_C
\sim
\left(\frac{\pi^2}{90}\right) g^{4/3}_{*,0}g^{-1/3}_{*,C} ~
\frac{T^4_0}{M_{\rm P}^2\,H_0^2} \, .
\end{equation}
Using the numerical values of $H_0$ and $T_0$, we find ${\cal D}_C \sim 10^{-5}$. 
It is seen that, as expected, $\mathcal{D}_A < \mathcal{D}_B < \mathcal{D}_C$. 

Based on the sequence of cooling and heating transitions, one would then expect a GW spectrum with two main features: 
\vskip 1.5mm
\noindent
{\bf (1) Multiple peaks}. The spectrum has a prominent peak (from $C$) that may be accompanied by a smaller discernible peak at higher frequency (due to $B$). The peak associated with $A$ is typically too suppressed to be observable.  
\vskip 1.5mm
\noindent
{\bf (2) Shallow tail}. 
Bubble collisions typically dominate GW produced from $B$ and their contribution, see~(\ref{eq:bc_spec_reform},\ref{sw}), falls off $\propto f^{-1}$ at $f \gg f^{\rm peak}_B$.  This leads to a distinct behavior in the high frequency tail of the spectrum. 
\vskip 1.5mm
\noindent
The total signal can then be used to probe the early thermal history in a straightforward way. One may directly read $T_{\rm PT}$ from  the main peak at $f^{\rm peak}_C$. The next peak at $f^{\rm peak}_B$, if visible, can be used to find $T_{\rm RH}$ through~(\ref{peaks}). In special cases, a third peak at $f^{\rm peak}_A$ associated with $A$ could be detected. This would allow us to infer $T_{\rm EQ}$ from Eq.~(\ref{peaks}). With both of $T_{\rm RH}$ and $T_{\rm EQ}$ known, one could use Eqs.~(\ref{tr2},\ref{R2}) to also determine the temperature at the onset of EMD $T_{\rm O}$.

\section{Results and Discussion}
\label{sec:result}
We now present our results for the GW spectrum from multiple FOPTs in the EMD scenario with non-monotonic evolution of temperature discussed in Section~\ref{sec:bfeqs}. First, we have numerically solved the set of Boltzmann equations to find evolution of temperature during the EMD epoch (see Appendix B for details). We have then computed contributions from the sound waves and bubble collisions for the cooling and heating transitions explained in Section \ref{sec:pt-gw}.  As far as the contribution of the sound  waves is concerned, we have included the recently improved results in the literature 
~\cite{Hindmarsh:2019phv,Guo:2020grp,Gowling:2021gcy,Guo:2024gmu}. In particular, we use an improved sound-shell fit based on Refs.~\cite{Guo:2024gmu,Banik:2025olw} that includes factors that capture the effect of equation of state of the background fluid from the spectral shape computed from hydrodynamics \cite{Guo:2024gmu} 
\begin{eqnarray}
\label{eq:sw_new_fit}
\Omega^{\rm GW}_{\rm sw,\,fit} h_{0}^2 &\simeq&
\frac{\mathcal{D}}{\mathcal{D}_{\rm std}}\,
\Omega_{\rm p} \left(\frac{f}{\tilde{f}_0}\right)^9
\frac{\,2 + \tilde{r}_b^{-12+\tilde{b}}\,}
{\left(\tfrac{f}{\tilde{f}_0}\right)^{\tilde{a}}
 + \left(\tfrac{f}{\tilde{f}_0}\right)^{\tilde{b}}
 + \tilde{r}_b^{-12+\tilde{b}}
   \left(\tfrac{f}{\tilde{f}_0}\right)^{12}}
\times \frac{\Upsilon(\omega)}{\Upsilon(1/3)} \left(\frac{1 + \alpha_*}{1+\alpha_* + R_{\phi,*}}\right)^2 \, . \nonumber \\
\,
\end{eqnarray}
Here $\mathcal{D}_{\rm std}$ is the amplitude redshift factor in the standard thermal history and $\Upsilon$ is the suppression ratio mentioned in Section \ref{sec:pt-gw}. The parameters $\Omega_{\rm p}$, $\tilde{f}_0$, $\tilde{a}$, $\tilde{b}$, and $\tilde{r}_b\!\equiv\! f_b/f_p$ control the peak height, characteristic scale, infrared and intermediate slopes, and the secondary-peak position respectively. They depend on $(\alpha_{\rm *},v_w,\beta/H_{\rm *},T_{\rm PT})$ as inputs in the code provided in Ref.~\cite{Guo:2024gmu}. Transitions $A$ and $B$ take place in the EMD epoch corresponding to $\omega = 0$ in Eq.~(\ref{eq:sw_new_fit}), while $C$ occurs in the RD phase for which $\omega = 1/3$.

We have chosen $6$ benchmark points shown in Table \ref{tab:params} that cover three separate regimes: $T_{\rm EQ} < T_{\rm PT} < T_{\rm RH}$ (the first two rows), $T_{\rm EQ} < T_{\rm PT} \sim T_{\rm RH}$ (the next three rows), and $T_{\rm EQ} \sim T_{\rm PT} < T_{\rm RH}$ (the last row). Two sets of PT parameters are shown for each benchmark point. The first set (numbers outside parentheses) employ equal values for all three transitions. The second set uses the same values for cooling transitions $A$ and $C$ but an alternative choice for the heating transition $B$. 
The corresponding GW spectra are shown  in Fig.~\ref{fig:hphi-temp-gw-tot}. The left panel in each row shows temperature evolution for a given benchmark point. The middle panel depicts the individual contributions to the GW spectrum (dashed) and the total spectrum (solid red) when all transitions have the same value of PT parameters. The right panel shows the individual contributions (dashed) and the total spectrum (solid gray) for alternative choice of the heating transition ($B$).   

We would like to emphasize that the three crossings of 
$T_{\rm EQ}$ in the left panels of Fig.~\ref{fig:hphi-temp-gw-tot} are not put in by hand; they 
follow from the numerical solution for $T(a)$. 
The relation $T_{\rm EQ} < T_{\rm PT} < T_{\rm RH}$ then makes the ordering of peak frequencies $f^{\rm peak}_C < f^{\rm peak}_B < f^{\rm peak}_A$ and the dilution pattern ${\cal D}_A < {\cal D}_{\rm B} < {\cal D}_C$ a direct consequence of the background evolution rather than a choice of plotting convention. Note that $m_\phi$ and $h$ can be swapped by $T_{\rm EQ}$ and $T_{\rm RH}$, see Eqs.~(\ref{tr2},\ref{R2}), for a given initial condition $H_{\rm O}$. We also note that $(T_{\rm EQ},T_{\rm RH},T_{\rm PT})$ are directly related to the GW observables. The results in Fig.~\ref{fig:hphi-temp-gw-tot} are therefore applicable to any EMD scenario that exhibits a similar non-monotonic evolution of $T$ regardless of the underlying particle physics model.    

Artificial intelligence (AI) has been used in the numerical analysis as an active search strategy for the FOPT parameter space. In this procedure, AI denotes an algorithmic loop in which an agent proposes candidate transition parameters, an emulator learns from previously evaluated spectra, and the full GW calculation is called only for selected candidates. The AI agent chooses sample points from the six dimensional vector $x=(v_w^{\rm mid},\alpha_{\rm PT}^{\rm mid},(\beta/H)_{\rm mid},v_w^B,$ 
\linebreak
$\alpha_{\rm PT}^B,r_\beta)$, where $r_\beta=(\beta/H)_B/(\beta/H)_{\rm mid}$, with common PT values for $A, B, C$ (the middle panels) and different values only for $B$ (the right panels). The emulator is a radial basis function, Gaussian process like surrogate that learns the expensive morphology score from exact spectrum evaluations and returns both a prediction $\mu(x)$ and an uncertainty estimate $\sigma(x)$. Then, the agent  uses an upper confidence bound acquisition rule, $\mathcal{A}(x)=\mu(x)+\kappa\sigma(x)$, to balance refinement near good candidates with exploration of uncertain regions. Candidate points that satisfy our criteria are accepted only if they respect the imposed physical and plotting constraints, including $v_w<0.99$, $\alpha_{\rm PT}<0.99$, $\beta/H<30$. In addition, they should follow the criteria on the allowed range of $r_\beta$, the required relic amplitude window for the total spectrum, the physical frequency ordering implied by the dilution and redshift factors, and the requirement that the right panel gray spectrum contains a visible secondary from the $B$ peak to the right of the main peak from $C$. 

The final benchmark values used in the plots are the output of this AI agent and emulator search that are reproducible. So that the selected $6$ benchmark points can be regenerated exactly while preserving the constraints from physics and the visual morphology of the spectra. The benchmark scan was restricted to points for which the hierarchy of peak locations, the dilution factors, and the detector windows are all satisfied at the same time. This is a useful criterion because it shows that the signal is not limited to a tuned point. Changing the transition strength, wall velocity, and inverse duration affect the relative height of the $B$ peak.  However, it does not remove the main mapping between the observed spectral shape and the temperatures $T_{\rm PT}$, $T_{\rm RH}$, and $T_{\rm EQ}$.

For all benchmark points the main peak from $C$ is clearly visible and within the reach of DECIGO, BBO, ET, or CE. The effect of the peak associated with $B$ is more visible for alternative choice of PT parameters, and may show up as a distinguishable second peak within the frequency range of ET or CE or broaden the main peak. In some cases, the shallow tail at high frequencies may also be detected by ET or CE.    

The detectability pattern also separates two questions that are often connected. The position of the dominant peak is controlled by $T_{\rm PT}$. However, the existence of the second peak and the shallow tail is governed by the amount of entropy injection and by the source type during the heating crossing. This separation causes the  identification of the cosmological origin of the signal even when the particle physics parameters of the transition are not known in advance.

\section{Summary and Conclusions}
\label{sec:conc-sum}
We have studied production of GW from multiple FOPTs in a non-standard cosmological history with non-monotonic evolution of temperature during an EMD epoch. Such a behavior can happen if the matter-like component that drives EMD has a time-dependent decay rate. Specifically, we considered a scenario where temperature increases in the non-adiabatic phase of EMD, $T \propto a^{3/8}$ as opposed to the standard relation $T \propto a^{-3/8}$. As explained, this can occur when a rotating complex scalar field (notably a flat direction in the scalar potential of SUSY extensions of the SM) dominates the energy density of the universe and drives an epoch of EMD.   

The most interesting situation arises when the PT temperature $T_{\rm PT}$ is lower than the reheating temperature at the end of the EMD $T_{\rm RH}$ and above the temperature at the beginning of the non-adiabatic phase of the EMD $T_{\rm EQ}$. In this case, three PTs take place in: a cooling transition in the adiabatic phase of EMD ($A$), a heating transition in the non-adiabatic phase of EMD ($B$), and a second cooling transition in the RD phase established after EMD ($C$). Sound waves are typically the main source of producing GW in $A$ and $C$, while bubble collisions can easily dominate GW production from $B$. 

Given the chronological order of the three PTs, the corresponding peak frequencies follow $f^{\rm peak}_C < f^{\rm peak}_B < f^{\rm peak}_A$. But $A$ also happens deep inside the EMD epoch resulting in an opposite order for the amplitude suppression factors $\mathcal{D}_A < \mathcal{D}_B < \mathcal{D}_C$. Thus, the overall GW spectrum has a prominent peak due to $C$ and a smaller peak at higher frequency from $B$. The peak associated with $A$ is usually too suppressed to show up (except in special cases). Moreover, in general, the high frequency tail of the spectrum is determined by the bubble collisions contribution from the heating transition $B$ resulting in a shallow fall off $\propto f^{-1}$. 

Our main results are shown in Fig.~\ref{fig:hphi-temp-gw-tot} where the middle and right panels depict the GW spectrum from multiple PTs for a few benchmark points. The primary peak from $C$ is always pronounced and within the detection range of  DECIGO, BBO, ET, and CE. In some cases, a second peak due to $B$ is discernible as a bump within the reach of  ET and CE. Frequency of the main peak is a proxy of $T_{\rm PT}$, and the second peak (if detectable) may be used to infer $T_{\rm RH}$ in the EMD scenario considered here. In special cases where a third peak (due to $A$) would be detectable, one could in addition determine $T_{\rm EQ}$ as well as temperature at the onset of EMD, $T_{\rm O}$, in this scenario. The high frequency tail of the spectrum may be within the frequency range of ET and CE, and its detection can be considered as an indication of an EMD scenario with non-monotonic temperature evolution. 

In summary, non-standard cosmological histories with an epoch of EMD may lead to multiple PTs in both cooling and heating phases. The resulting GW spectrum exhibits features 
that can be tested by the upcoming and future GW detectors and thereby allow us to probe the early thermal history.     

\acknowledgments
The work of R.~A. is supported in part by NSF Grant No. PHY-2210367. F.~H. is supported by Homer Dodge postdoctoral fellowship. F.H. would like to thank Amitayus Banik, Nicol{\'a}s Bernal, Dietrich B{\"o}deker, Kuver Sinha, Graham White, and Yang Xiao for useful discussions. The authors thank the George P. and Cynthia Woods and Mitchell Institute for Fundamental Physics and Astronomy (MIFP) at Texas A$\&$M University, where this work began, for their kind hospitality and support. 
They are also grateful to the Center for Theoretical Underground Physics and Related Areas (CETUP*), the Institute for Underground Science at Sanford Underground Research Facility (SURF), and the South Dakota Science and Technology Authority for their hospitality and financial support at different stages of this work. In this paper, we have used AI tools in polishing the text and improving the codes for the plots.

\appendix
\section{Loop-level Decay to Gauge Bosons}
\label{sec:appA}

Consider a real scalar field $\phi$ with a Yukawa coupling to a Dirac fermion $\psi$ that is charged under some gauge group
\begin{equation} 
\label{lagran4}
  {\cal L} \supset h \phi {\bar \psi} \psi + {\bar \psi}  \gamma^\mu D_\mu \psi .
\end{equation}
These interactions induce an effective operator that couples $\phi$ to gauge bosons via a one-loop triangle diagram (see Fig.~\ref{fig:phi_decay}). For a non-abelian group with gauge coupling $g$, we have~\cite{Djouadi:2005gj,Spira:2016ztx}
\begin{eqnarray}
\mathcal{L}_{\rm eff}
&\supset&
c_{VV}\,\phi\,F^a_{\mu\nu} F^{a\mu\nu},
\label{eq:effop}
\end{eqnarray}
where
\begin{eqnarray}
\label{eq:couplingvv}
c_{VV}
&=&
\frac{g^2\,h}{16\pi^2\,m_{\psi}}\;N_f\,C(r)\;A_{1/2}(\tau) \quad ,
\quad
\tau \equiv \frac{4m_{\psi}^2}{m_{\phi}^2} .
\end{eqnarray}
Here $C(r)$ is the Dynkin index  of the fermion representation ($C(r)=\tfrac12$ for fermions in the fundamental representation of  SU($N$)), and $N_f$ is the number of fermions running in the loop. The loop function is given by
\begin{eqnarray} 
\label{eq:a12}
A_{1/2}(\tau)
=
2\tau\bigl[1+(1-\tau)f(\tau)\bigr],
\end{eqnarray}
where
\begin{eqnarray}
\label{eq:ftau}
f(\tau)
=
\begin{cases}
\arcsin^2(1/\sqrt{\tau}),&\tau\ge1,\\[4pt]
-\tfrac14\left[\ln\frac{1+\sqrt{1-\tau}}{1-\sqrt{1-\tau}}-i\pi\right]^2,&\tau<1.
\end{cases}
\end{eqnarray}
The resulting partial decay width into gauge bosons $V$ is given by~\footnote{For a complex scalar $\phi$, the expression in Eq.~(\ref{lagran3}) must be replaced with that in~(\ref{lagran2}). In this case, the real and imaginary components of $\phi$ are coupled to the gauge bosons via the effective operators $\phi_{\rm R} F F$ and  $\phi_{\rm I} F {\tilde F}$ respectively, leading to the the same $\Gamma^{\rm loop}_\phi$ as in~(\ref{eq:gamma-v}). }
\begin{eqnarray}
\label{eq:gamma-v}
\Gamma_\phi^{\rm loop}  \equiv \Gamma(\phi\to VV) &=&
\frac{|c_{VV}|^2\,m_{\phi}^3}{8\pi}\,. 
\end{eqnarray}
For an abelian gauge group, $g^2 C(r)$ should be replaced with $ \sum_{f}{Q^2_f}$ where $Q_f$ denotes the $U(1)$ charge of fermions in the loop.  

We can find the asymptotic behavior of $\Gamma^{\rm loop}_\phi$ for small and large values of $\tau$ from Eqs. (\ref{eq:a12},\ref{eq:ftau}):
\begin{eqnarray} 
\label{eq:gammas}
&& \Gamma^{\rm loop}_\phi \propto {1 \over m^2_\psi} \qquad \qquad \tau \gg 1 \, , \\
&& \Gamma^{\rm loop}_\phi \rightarrow 0 \qquad \qquad \quad\tau \ll 1 \, .
\end{eqnarray}

To put things in perspective, consider the case where $\phi$ represents the $H_u H_d$ flat direction in SUSY extensions of the SM. In this case, Yukawa terms for quarks and leptons induce one-loop diagrams that couple $\phi$ to gluons and the photon as described above.

\section{Numerical Solutions of Boltzmann-Friedmann Equations}
\label{app:boltz-fried}

In this appendix, we describe the methodology for numerically solving the set of Boltzmann-Friedmann equations that controls the evolution of respective energy densities. The scalar field energy density $\rho_{\phi}$ and radiation energy density ${\rho}_{\rm R}$ evolve according to \cite{Drees:2017iod,Drees:2018dsj,Bernal:2019lpc,DEramo:2019tit}.
\begin{eqnarray}
&& \frac{d\rho_{\phi}}{dt} + 3H \rho_{\phi} =  - \Gamma_{\phi} \rho_{\phi},
\\
&& \frac{d\rho_{\rm R}}{dt} + 4H \rho_{\rm R}  =   + \Gamma_{\phi} \rho_{\phi},
\\
\label{eq:bolfried3}
\end{eqnarray}
where $H^2 = (\rho_\phi + \rho_{\rm R})/3 M^2_{\rm P}$. The total decay width of $\phi$ includes contributions from the tree-level and loop-level channels
\begin{equation}
\Gamma_{\phi}  \equiv  \Gamma_\phi^{\rm loop} \Theta[h {\tilde \phi}(t) - m_{\phi}/2] + \Gamma^{\rm tree}_{\phi} \Theta[-h {\tilde \phi}(t) + m_{\phi}/2]\,,
\end{equation}
where ${\tilde \phi} \equiv (\phi_{\rm R}^2 + \phi_{\rm I}^2)^{1/2}$. For a circularly rotating scalar field it is related to the field energy density according to $\rho_\phi = m^2_\phi {\tilde \phi}^2$. 

The Heaviside function $\Theta[h {\tilde \phi}(t) - m_{\phi}/2]$ introduces a conditional dependence on the field amplitude that represents the threshold for the loop-level decay channel to turn on (it equals unity for $h{\tilde \phi} > m_\phi/2$, when tree-level decay is kinematically blocked). The second Heaviside function $\Theta[-h {\tilde \phi} + m_\phi/2]$ is used as an approximation to underscore the fact that $\Gamma^{\rm loop}_\phi \rightarrow 0$ for $h {\tilde \phi} \ll m_\phi$. In fact, see Eqs.~(\ref{loopdec},\ref{eq:treedec}), $\Gamma^{\rm tree}_\phi$ totally dominates over $\Gamma^{\rm loop}_\phi$ for $h {\tilde \phi} \lesssim m_\phi/2$.  

Making a change of variables $t\to a$, and using the comoving energy density $\Phi = \rho_{\phi} a^3$, 
we find
\begin{eqnarray}
\label{eq:comoving}
&& \frac{d\Phi}{da} = -\frac{\Gamma_{\phi}}{a\,H}\,\Phi,\\
&& \frac{dT}{da} 
= \Bigl(-\frac{T}{a} + \frac{\Gamma_{\phi}\,\Phi}{3 a^4\,H\,s }\Bigr) \Bigl[1 + \frac{T}{3g_{*, \rm S}}
  \frac{dg_{*, \rm S}}{dT}\Bigr]^{-1} \, ,
\end{eqnarray}
where $s = 2\pi^2 g_{*S}T^3/45$ is the entropy density. The initial conditions for $\Phi$ and $T$ are related to those for the energy densities as follows
\begin{eqnarray}
\label{eq:init1}
\rho_{{\rm R},{\rm O}} = \frac{\pi^2}{30} g_{*,{\rm O}}T_{\rm O}^4 \qquad,   \qquad\rho_{\phi,{\rm O}} = m^2_{\phi}{\tilde \phi}_{\rm O}^2 \, . \,  \, 
\end{eqnarray}
Then, defining the onset of EMD epoch as the moment where  $\rho_{\rm R} = \rho_\phi$, the value of $\phi_{\rm O}$ is set in terms of $T_{\rm O}$ as follows
\begin{eqnarray}
   {\tilde \phi}_{\rm O} = \left [\frac{\pi^2}{30} \frac{g_{*,{\rm O}}T_{\rm O}^4}{m^2_{\phi}}\right]^{\frac{1}{2}}\,. 
\end{eqnarray}

Using the expressions in Eqs.~(\ref{loopdec},\ref{eq:treedec}) for $\Gamma^{\rm loop}_\phi$ and $\Gamma^{\rm tree}_\phi$ respectively, with $g \approx 0.3$ and $C \approx 1$, we have integrated the coupled differential equations in~(\ref{eq:comoving}) to obtain $T(a)$ for various values of $m_\phi$ and $h$ scanned over the following range in the $m_\phi-h$ plane
\begin{eqnarray}
\label{eq:init2}
10^{-6} \leq h \leq 1 \qquad \qquad,\qquad \qquad 10^5 ~ {\rm GeV} \leq m_\phi \leq 10^{10} ~ {\rm GeV} \,. 
\end{eqnarray}
We note that the range chosen for $h$ is in agreement with the SM Yukawa couplings. In addition, the minimum value of $m_\phi$ in~(\ref{eq:init2}) satisfies 
the condition $m_{\phi} > H_{\rm O}$ that is required for the $\phi$ field to behave as a matter-like component.  

\begin{figure}[htbp!]
    \centering
\includegraphics[width=0.4\textwidth]{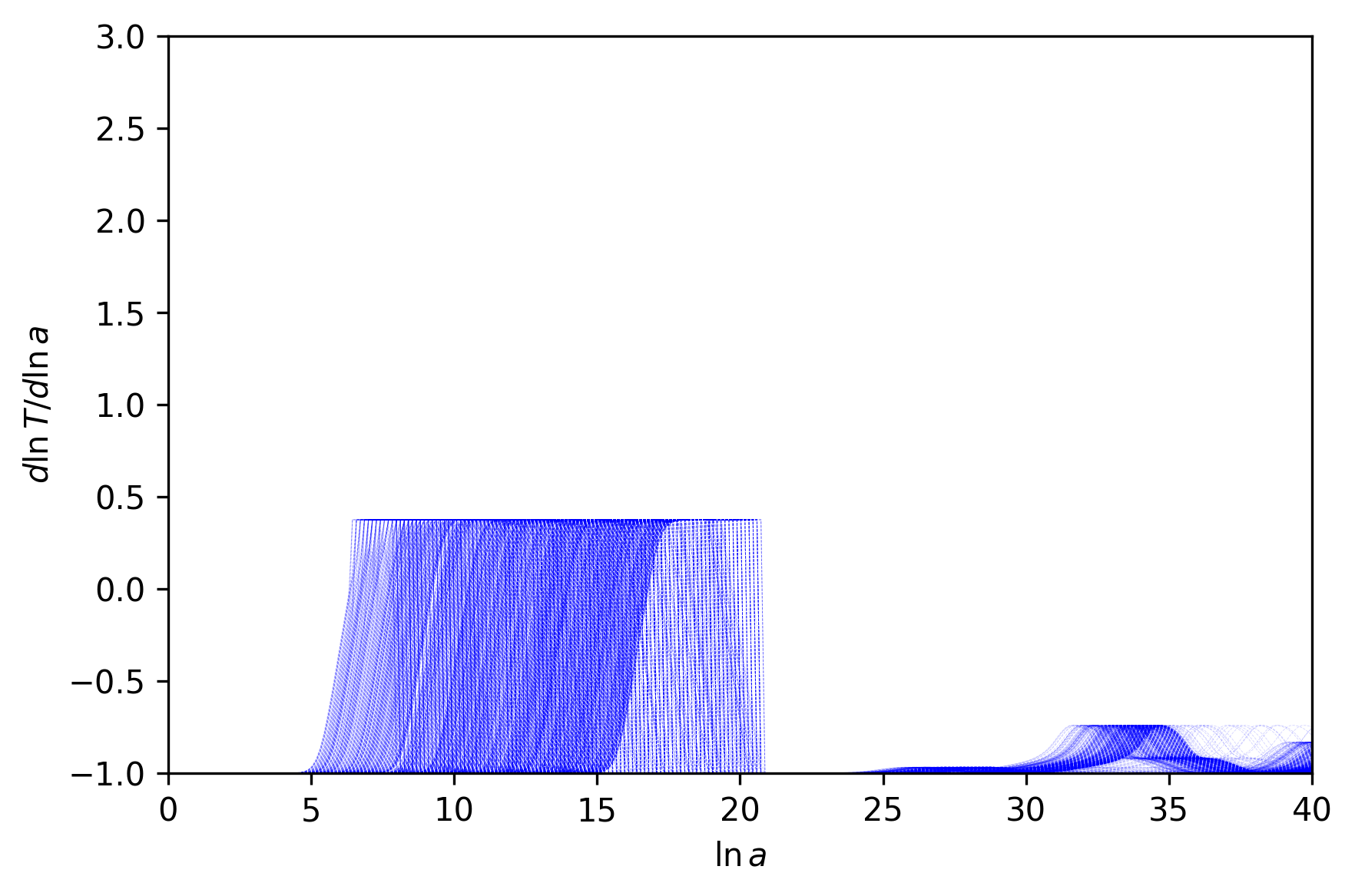}
\includegraphics[width=0.4\textwidth]{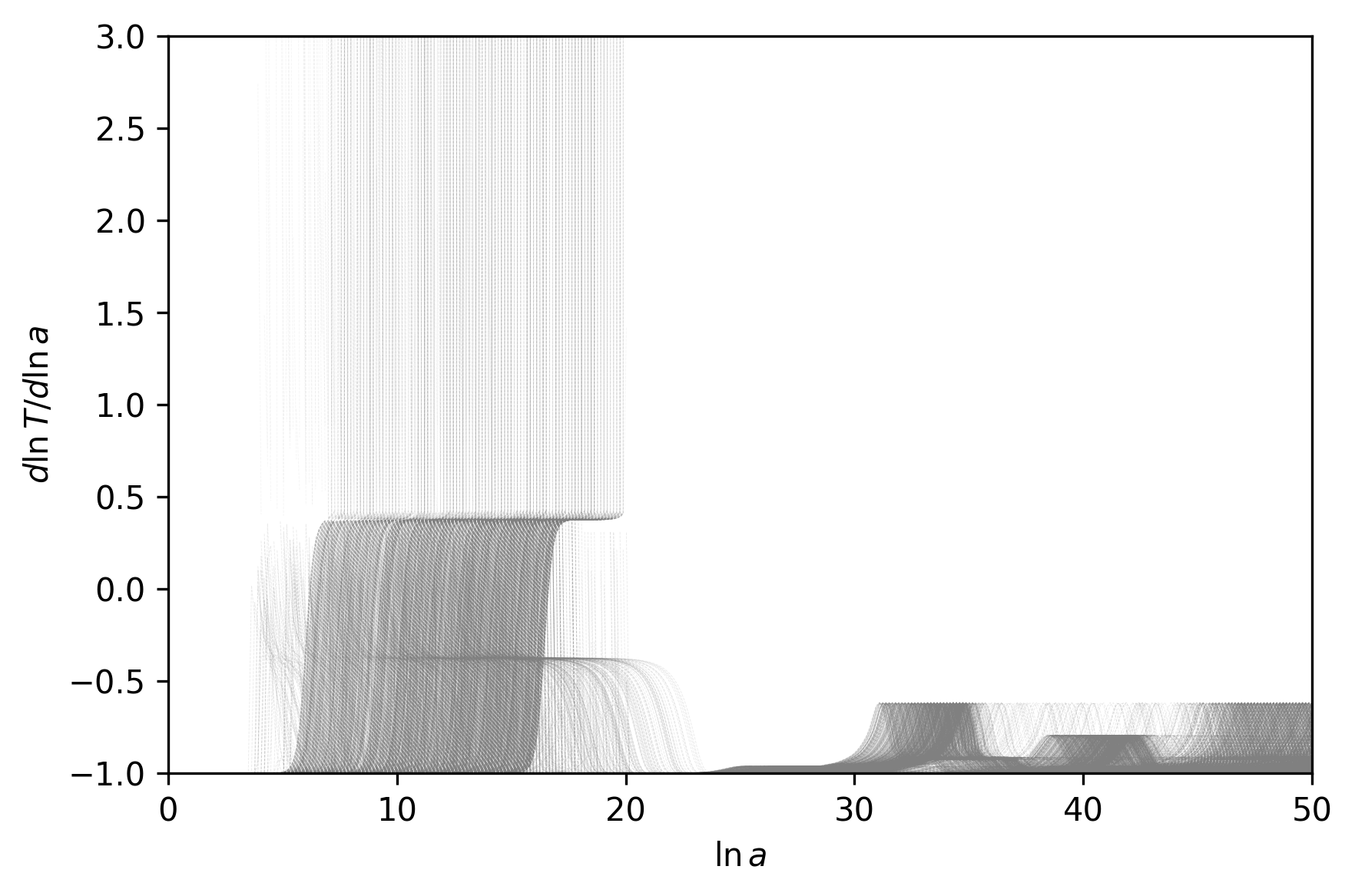}
\includegraphics[width=0.4\textwidth]{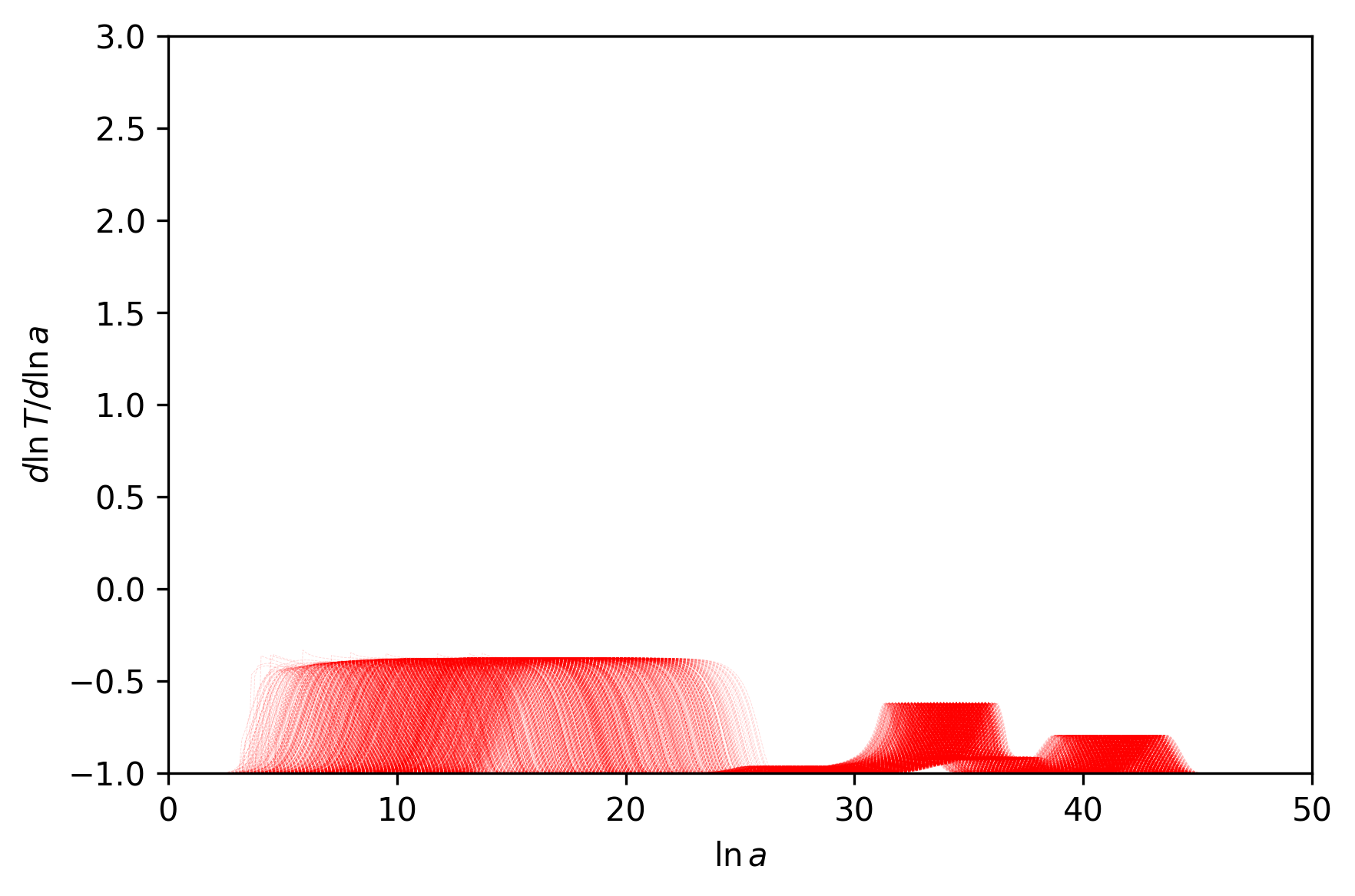}
    \caption{Temperature of the universe as a function of the scale factor for the points shown in Fig.~2. All cases start in  the adiabatic phase of EMD where $T \propto a^{-1}$. In the upper left panel, this is followed by $T \propto a^{3/8}$ in the non-adiabatic that smoothly switches back to  $T \propto a^{-1}$ at the end of EMD. In the upper right  panel, $T$ shows a similar behavior but with a large spike at the end of EMD. The lower panel shows the standard evolution where $T \propto a^{-3/8}$ during the non-adiabatic phase of EMD. 
    }
    \label{fig:m-h2}
\end{figure}

Fig.~\ref{fig:m-h2} depicts evolution of $T$ as a function of $a$ (in units of its initial value $a_{\rm O}$) for the scanned points in the $m_\phi-h$ plane. We can recognize three main groups in this figure shown in blue, gray, and red:
\vskip 1.5mm
\noindent{\bf (1)} The group in blue (lower left panel) represents all points for which $T\propto a^{-1}$ in the adiabatic phase of EMD followed by $T \propto a^{3/8}$ during the non-adiabatic phase and then switched to $T \propto a^{-1}$ in the RD phase. In some cases, transition to RD is accompanied by a small increase in temperature at the end of EMD. This happens when the tree-level decay channel becomes kinematically open around the time when $H \sim \Gamma^{\rm loop}_\phi$. Then, given that $\Gamma^{\rm tree}_\phi \gg \Gamma^{\rm loop}_\phi$, the energy density of $\phi$ decays at a much faster rate at the very end of EMD. This results in an increase in $T$, albeit by an ${\cal O}(1)$ factor, given that a significant fraction of $\rho_\phi$ has already decayed by this time.     
\vskip 1.5mm
\noindent{\bf (2)} The group in gray represents (upper right panel) the points for which the $T \propto a^{3/8}$ behavior in the non-adiabatic phase of EMD is accompanied by a large increase  in temperature at the end of EMD epoch. This happens when the tree-level decay channel is turned on earlier when $\Gamma^{\rm loop}_\phi \ll H \lesssim \Gamma^{\rm tree}_\phi$. In this case, only a small fraction of $\rho_\phi$ has decayed when the tree-level channel opens up. The bulk of $\phi$ energy density then decays very quickly (compared to the Hubble time) resulting in a sharp rise in $T$ at transition to RD.    
\vskip 1.5mm
\noindent{\bf (3)} The group in red (lower panel) represents all points for which $T\propto a^{-3/8}$ in the non-adiabatic phase of EMD. This is because the tree-level decay channel is open during the adiabatic phase of EMD (in some cases at its very onset) leading to the standard scenario with monotonically decreasing temperature throughout EMD.
\vskip 1.5mm
Fig.~\ref{fig:m-h} in the text shows these three groups in the $m_\phi-h$ plane. The red group is not of interest to us because it gives rise to the standard EMD scenario. The blue group is the most interesting of all as it represents a case where the loop-level decay leads to non-monotonic behavior of temperature and is also responsible for termination of the EMD epoch and a smooth transition to RD.      

\begin{figure}[t]
\centering

\begin{tabular}{cc}
\includegraphics[width=0.47\textwidth]{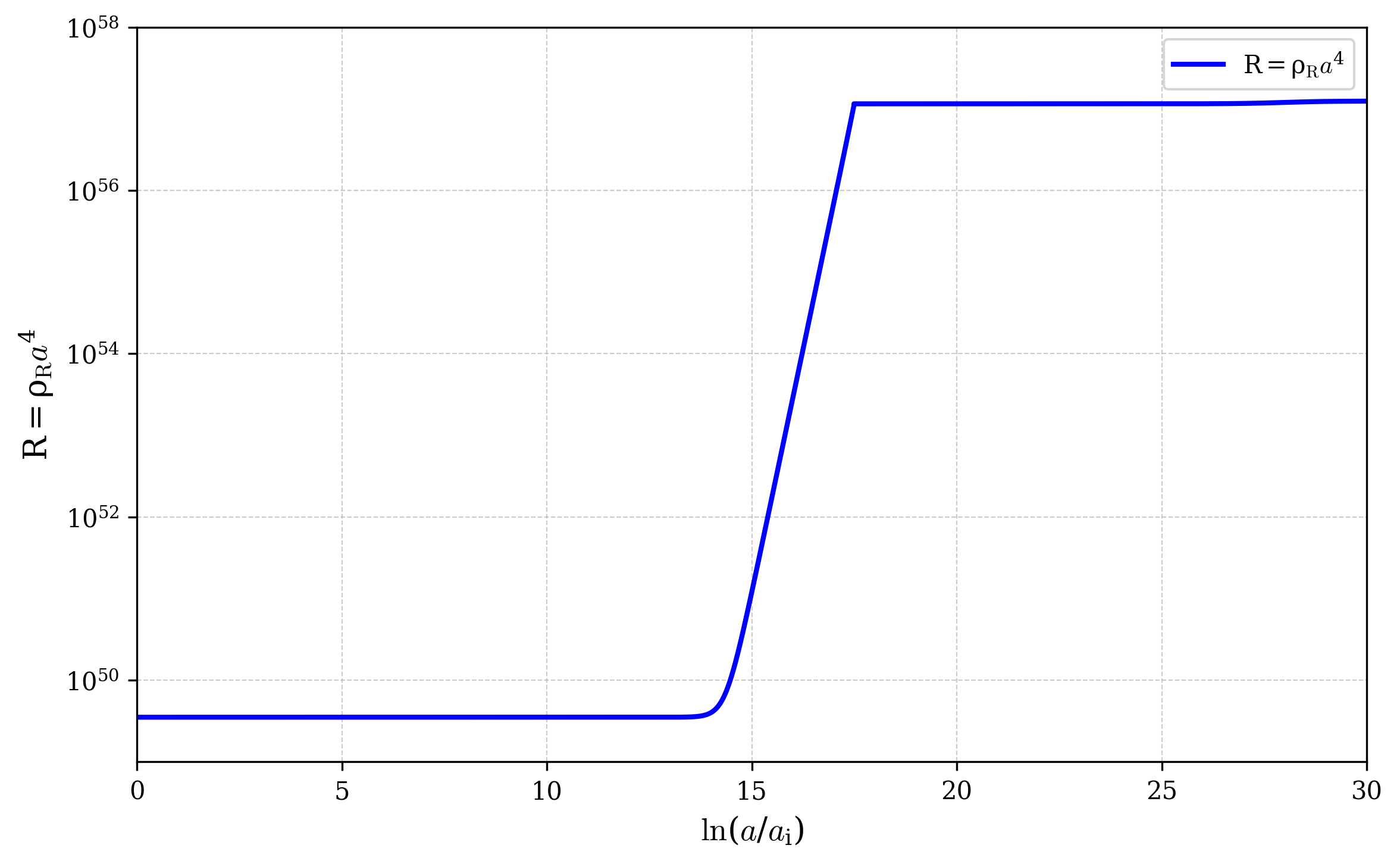}
&
\includegraphics[width=0.47\textwidth]{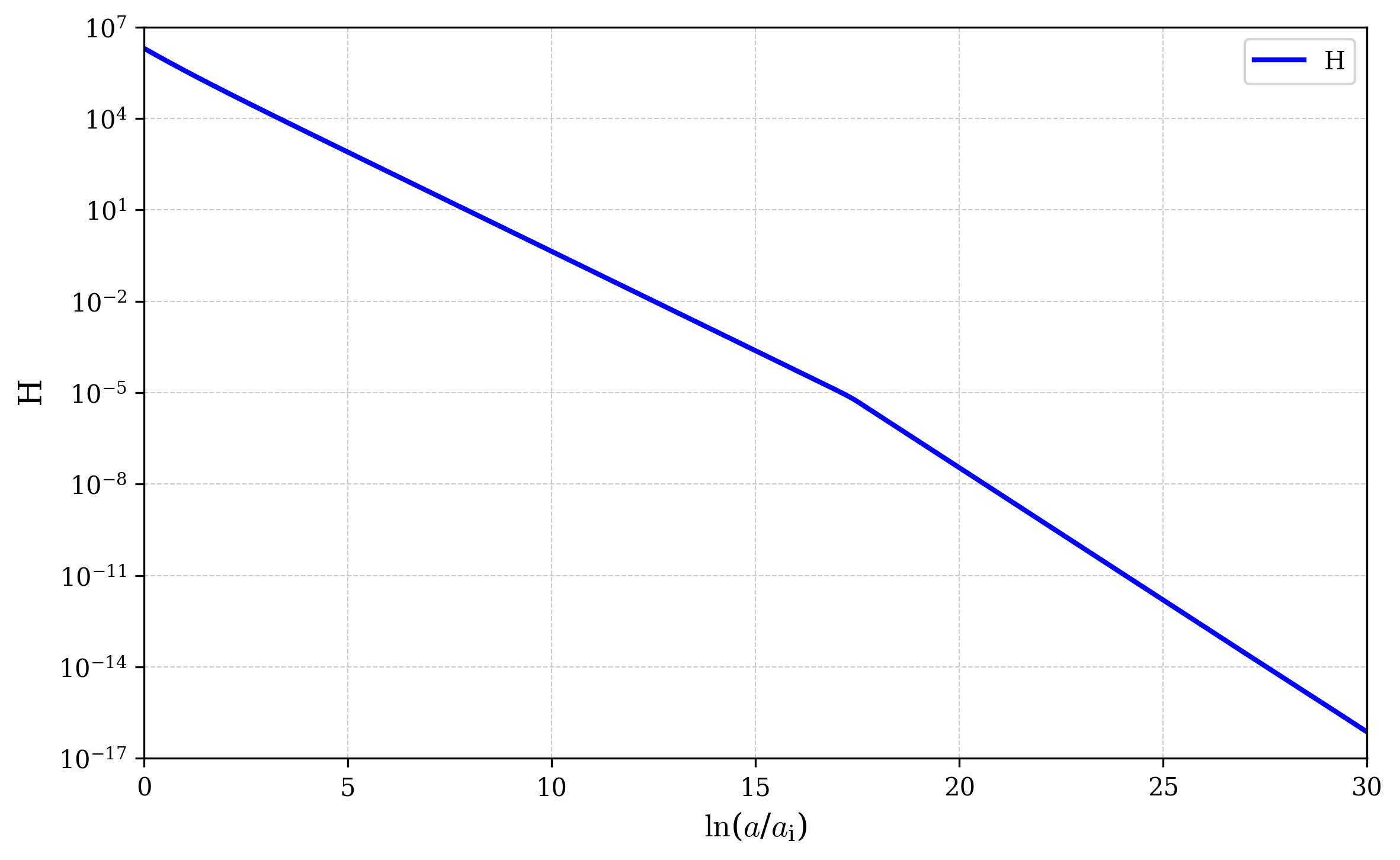}
\\[1.0em]

\includegraphics[width=0.47\textwidth]{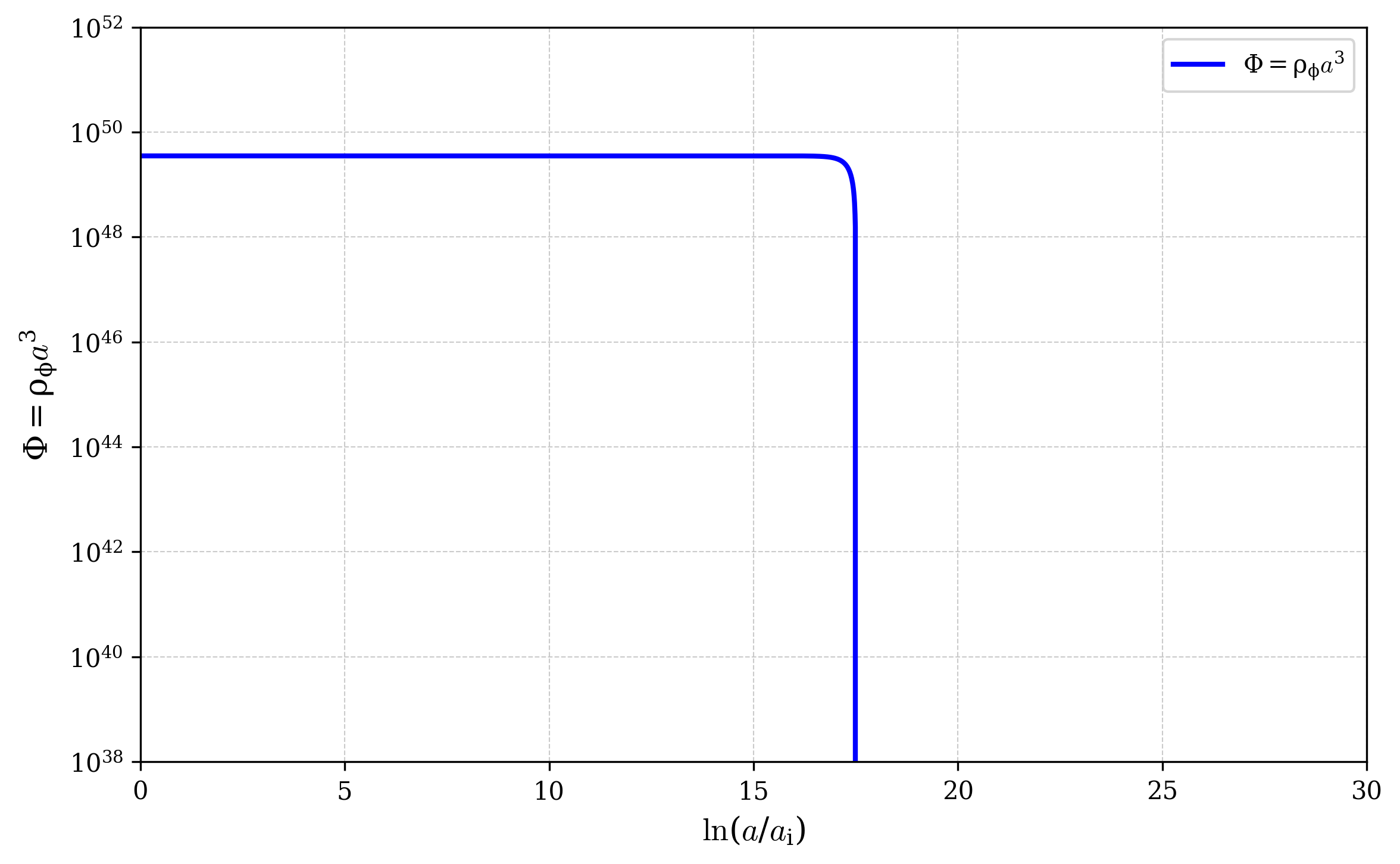}
&
\includegraphics[width=0.47\textwidth]{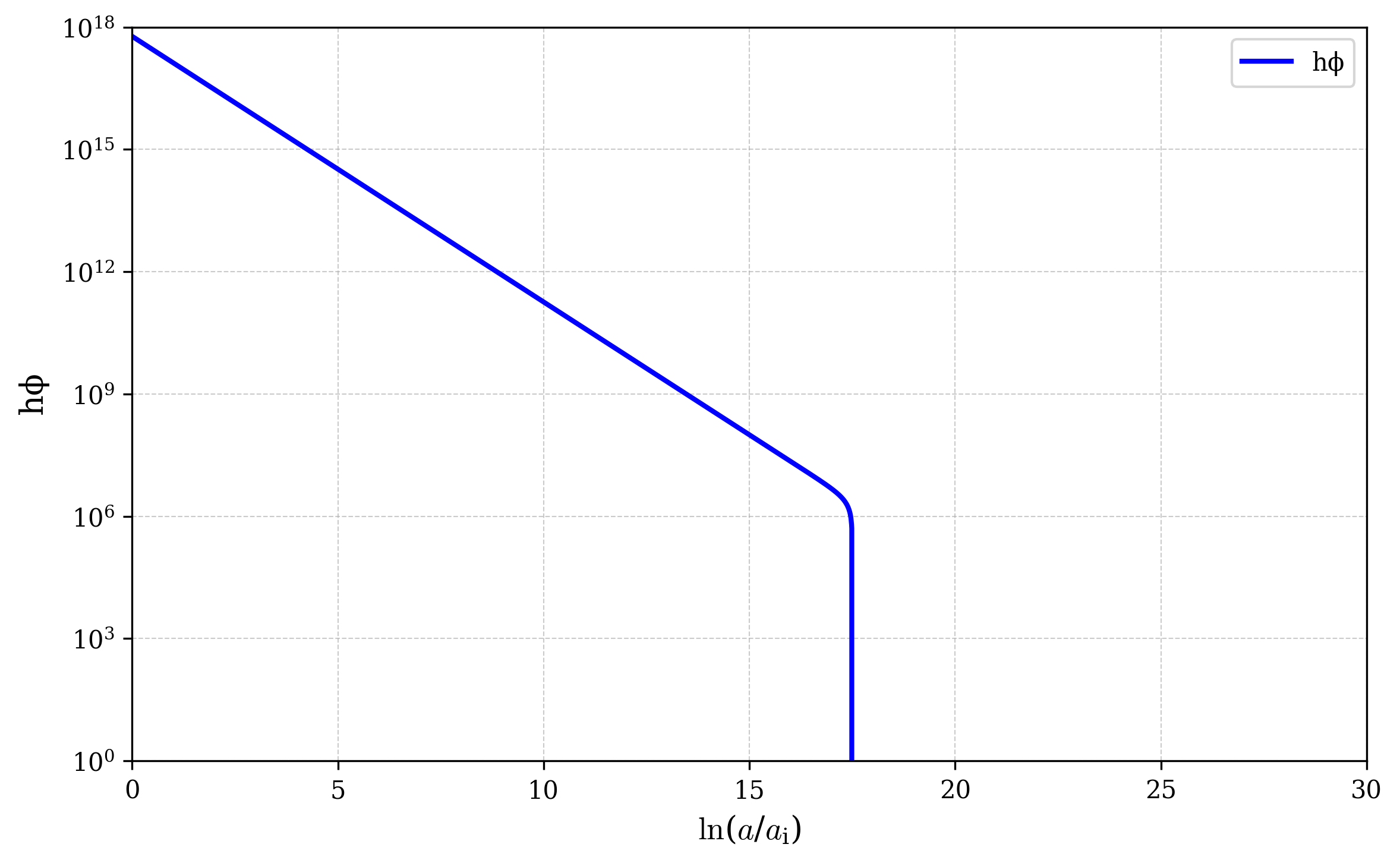}
\end{tabular}
    \caption{Evolution of various quantities as a function of the scale factor (expressed in units of its initial value) for the benchmark point with $m_\phi = 10^6$ GeV and $h = 10^{-1}$ and the initial condition $T_{\rm O} = 10^{11}$ GeV. Different panels show the comoving radiation energy density (top left panel), the Hubble expansion rate (top right panel), the comoving scalar energy density (bottom left panel), and $h{\tilde \phi}$ (bottom right panel). }
    \label{fig:comov}
\end{figure}

\section{Peak Frequencies and Amplitudes of Gravitational Waves from Multiple Transitions}
\label{app:peak-ratio}

Here we compare the peak frequencies and amplitudes of different components of the GW produced from multiple transitions at the same temperature $T_{\rm PT}$. As mentioned, the first peak is due to the cooling transition during the adiabatic phase of EMD occurring at $H_{\rm A}$. The second one comes from the heating transition in the non-adiabatic phase of EMD that occurs at $H_{\rm B}$. The third peak is from the second cooling transition occurring in the RD phase at $H_{\rm C}$.   

For a peak generated at $(H_*,a_*)$, the observed frequency at the present time (denoted by subscript ``0') is given by $f \sim H_{*} (a_{*}/a_0)$. Therefore, taking $a_0 = 1$, we have
\begin{eqnarray}
f^{\rm peak}_{A} &\sim& H_{A} a_{A} \, , \nonumber \\
f^{\rm peak}_{B} &\sim& H_{B} a_{B} \, , \nonumber \\
f^{\rm peak}_{C} &\sim& H_{C} a_{C} \, , \nonumber \\
\end{eqnarray}
Then, given that $H \propto a^{-2}$ during RD, we find
\begin{equation} 
\label{eq:eq1}
f^{\rm peak}_C  \sim H_{\rm RH} a_{\rm RH}~ \left({H_{\rm *} \over H_{\rm RH}}\right)^{1/2} \sim H_{\rm RH} a_{\rm RH}\left({g_{*,C} \over g_{*,{\rm RH}}}\right)^{1/4} ~ {T_{\rm PT} \over T_{\rm RH}} , 
\end{equation}
where we have used the relation between $H$ and $T$ during RD in the last step. On the other hand, $H \propto a^{-3/2}$ during EMD . After using the relation between $H$ and $T$ in the non-adiabatic phase of EMD, see Eq.~(\ref{nonad2}), we find
\begin{equation} 
\label{eq:eq2}
f^{\rm peak}_B \sim H_{\rm RH} a_{\rm RH} \left({H_B \over H_{\rm RH}}\right)^{1/3} \sim H_{\rm RH} a_{\rm RH} \left({g_{*,{\rm RH}} \over g_{*,B}}\right)^{1/3} \left({T_{\rm RH} \over T_{\rm PT}}\right)^{4/3}.
\end{equation}
For comparable values of $\beta/H_{\rm *}$ at $B$ and $C$, and noting that $g_*$ remains essentially the same at high temperatures between $T_{\rm PT}$ and $T_{\rm RH}$, we then find
\begin{equation} \label{peak1}
f^{\rm peak}_{B} \sim \left({T_{\rm RH} \over T_{\rm PT}}\right)^{7/3} f^{\rm peak}_C.
\end{equation}
Similarly
\begin{equation}
{f^{\rm peak}_{A} \over f^{\rm peak}_{B}} \sim {H_A a_{A} \over H_{B} a_{B}}    
\end{equation}
Given that $H \propto a^{-3/2}$ during EMD, this results in
\begin{equation}
\label{fABrat}
{f^{\rm peak}_{A} \over f^{\rm peak}_{B}} \sim \left({H_A \over H_B}\right)^{1/3} \,.
\end{equation}
Note that
\begin{equation}
 {H_A \over H_B} = {H_A \over H_{\rm EQ}} {H_{\rm EQ} \over H_B} \simeq \left({g_{*,A} \over g_{*,\rm EQ}}\right)^{3/8} \left({T_{\rm PT} \over T_{\rm EQ}}\right)^{3/2} \left({g_{*,B} \over g_{*,{\rm EQ}}}\right) \left({T_{\rm PT} \over T_{\rm EQ}}\right)^4 ,   
\end{equation}
Then, after using Eqs.~(\ref{ad},\ref{nonad2}), we find 
\begin{equation} 
\label{peak2}
f^{\rm peak}_A \sim \left({T_{\rm PT} \over T_{\rm EQ}}\right)^{11/6} f^{\rm peak}_B \,.  
\end{equation}

As for the suppression factor, we have $\mathcal{D} = (a_*/a_0)^4 (H_0/H_*)^2$. Then, since $a_0 = 1$, we have
\begin{eqnarray}
&& \mathcal{D}_B = \left({a_B \over a_C}\right)^4 \left({H_B \over H_C}\right)^2 \mathcal{D}_C \, ,\\
\end{eqnarray}
We note that $(a_{\rm RH}/a_C)^2 = (H_C/H_{\rm RH})$ and $(a_{\rm RH}/a_{A,B}) = (H_{A,B}/H_{\rm RH})^{2/3}$ . Thus
\begin{eqnarray}
&& \mathcal{D}_{A,B} = \left({H_{\rm RH} \over H_{A,B}} \right)^{2/3} \mathcal{D}_C \, . \\
\end{eqnarray}
After using Eqs.~(\ref{ad},\ref{nonad2}), and noting that $g_{*}$ essentially remains constant during EMD, we find
\begin{eqnarray} \label{suppression1}
 && \mathcal{D}_B \sim \left({T_{\rm PT} \over T_{\rm RH}}\right)^{8/3} \mathcal{D}_C\, ,
\end{eqnarray}
and 
\begin{eqnarray} \label{suppression2}
\mathcal{D}_A \sim \left({T_{\rm EQ} \over T_{\rm PT}}\right)^{11/3} \mathcal{D}_B
\, .
\end{eqnarray}

\bibliography{biblio.bib}
\end{document}